\documentclass[oneside,english]{article}
\usepackage[T1]{fontenc}
\usepackage[latin1]{inputenc}
\usepackage{array}
\usepackage{float}
\usepackage{amsmath}
\usepackage{graphicx}
\usepackage{geometry}
\usepackage{amssymb}
\usepackage[numbers]{natbib}

\makeatletter

\providecommand{\tabularnewline}{\\}

\usepackage{babel}

\ifx\pdfoutput\@undefined\usepackage[usenames,dvips]{color}
\else\usepackage[usenames,dvipsnames]{color} \fi
\usepackage{hyperref}

\makeatother
\begin{document}

\title{\begin{flushright}\textit{\large in press in Int. J. Mass Spec.\\~\\}\end{flushright}%
The WITCH experiment: completion of a set-up to investigate the
structure of weak interactions with a Penning trap.}
\author{V.Yu.~Kozlov$^{\,a),\,}$\footnote{\,Corresponding author: %
\href{mailto:Valentin.Kozlov@fys.kuleuven.ac.be}{Valentin.Kozlov@fys.kuleuven.ac.be}}\,\;, %
N.~Severijns$^{\,a)}$, D.~Beck$^{\,b)}$, M.~Beck$^{\,a)}$, S.~Coeck$^{\,a)}$,\\
B.~Delaur{\'e}$^{\,a)}$, S.~Kopecky$^{\,a)}$, A.~Lindroth$^{\,a)}$, P.~Delahaye$^{\,c)}$,\\
F.~Wenander$^{\,c)}$, V.V.~Golovko$^{\,a)}$, I.S.~Kraev$^{\,a)}$, T.~Phalet$^{\,a)}$, \\
(ISOLDE, NIPNET and TRAPSPEC collaborations)\\~\\%
{\small $^{\,a)}$~Instituut voor Kern- en Stralingsfysica,
K.U.~Leuven,}\\
{\small Celestijnenlaan~200D, B-3001~Leuven, Belgium}\\
{\small $^{\,b)}$~GSI-Darmstadt, Planckstr.~1, D-64291~Darmstadt,
Germany}\\
{\small $^{\,c)}~$CERN, CH-1211~Gen\`eve 23, Switzerland}}

\maketitle

\begin{abstract} The WITCH experiment aims to study a possible admixture of a scalar or tensor type interaction in $\beta$-decay by determining the  $\beta-\nu$ angular correlation from the shape of the recoil energy spectrum. The installation period was completed and intensive commissioning of the set-up was performed already. The lay-out of the WITCH set-up and results of commissioning tests performed until now are described here, showing that the full set-up up to the spectrometer is now operational, although several efficiencies are still to be improved. Due to its feature of being able to measure the energy spectrum for recoil ions, the WITCH experiment also opens possibilities for other
observables. 
\end{abstract}

\mark{{}{in press in Int. Jour. Mass Spec.}}

\section{Introduction}

Despite the fact that the $\beta$-decay process was discovered
already at the end of the 19th century, our understanding of weak
interactions has developed only gradually. The Standard Model of
the electroweak interaction is very successful in describing
existing experimental data both qualitatively and quantitatively,
but a number of parameters have to be determined experimentally
and several important properties of the interaction are not well
understood. The most general interaction Hamiltonian for nuclear
$\beta$-decay which includes all possible interaction types
consistent with Lorentz-invariance is given
by~\cite{lee56,jackson57a}

\begin{eqnarray}
\mathcal{H}_{\beta} & = & \;\left(\bar{p}n\right)\left(\bar{e}\left(C_{S}+C_{S}^{\prime}\gamma_{5}\right)\nu\right)\nonumber \\
 & + & \:\left(\bar{p}\gamma_{\mu}n\right)\left(\bar{e}\gamma^{\mu}\left(C_{V}+C_{V}^{\prime}\gamma_{5}\right)\nu\right)\nonumber \\
 & + & \:\frac{1}{2}\left(\bar{p}\sigma_{\lambda\mu}n\right)\left(\bar{e}\sigma^{\lambda\mu}\left(C_{T}+C_{T}^{\prime}\gamma_{5}\right)\nu\right)\nonumber \\
 & - & \:\left(\bar{p}\gamma_{\mu}\gamma_{5}n\right)\left(\bar{e}\gamma^{\mu}\gamma_{5}\left(C_{A}+C_{A}^{\prime}\gamma_{5}\right)\nu\right)\nonumber \\
 & + & \:\left(\bar{p}\gamma_{5}n\right)\left(\bar{e}\gamma_{5}\left(C_{P}+C_{P}^{\prime}\gamma_{5}\right)\nu\right)+\: h.c.\label{eq:H-general}\end{eqnarray}

\noindent The coefficients $C_{i}$ and $C_{i}^{\prime}$, $i\in\{
S,V,T,A,P\}$ are the coupling constants for Scalar (S), Vector
(V), Tensor (T), Axial-Vector (A) and Pseudoscalar (P)
contributions. In the Standard Model of the weak interaction only
V and A interactions are present at a fundamental level, which
leads to the well-known $V-A$ structure of the weak interaction.
However, this assumption is based on experimental results only and
the presence of scalar and tensor types of weak interaction is
today ruled out only to the level of about 8\% of the V- and
A-interactions~\cite{severijns05a}.

\begin{figure}[H]
\centering

\includegraphics[%
  width=0.99\textwidth,
  keepaspectratio]{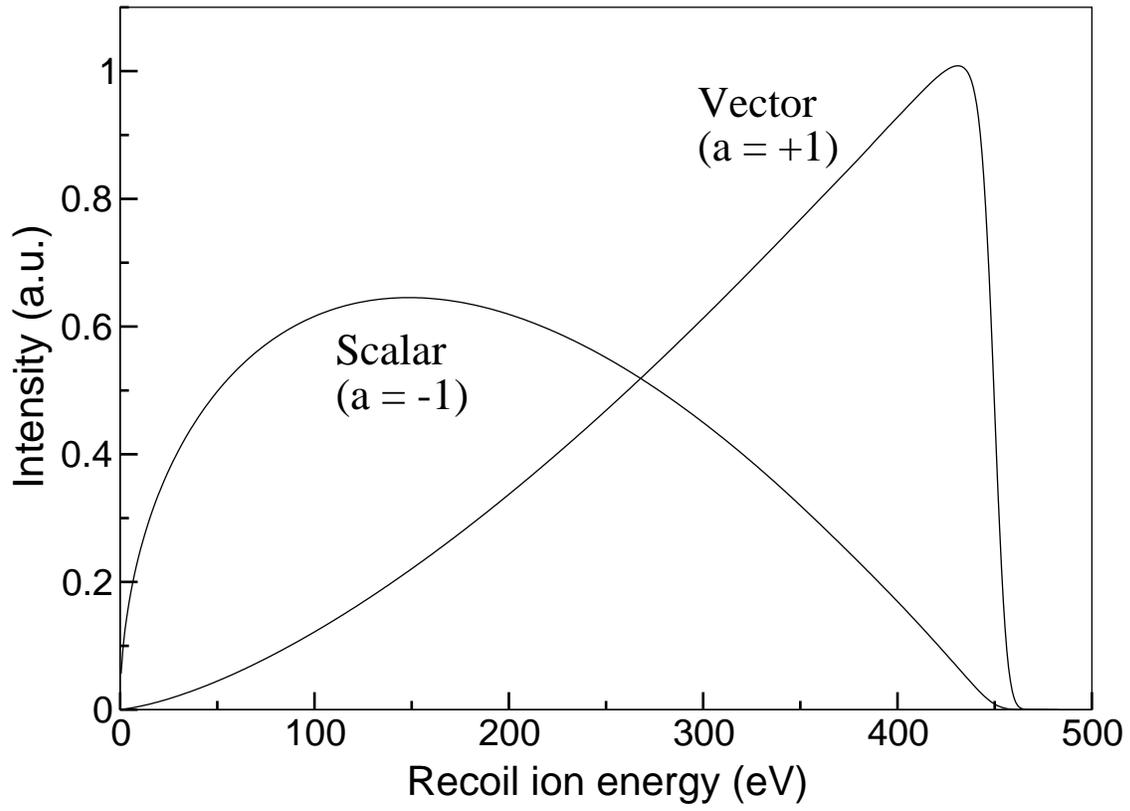}

\caption{\label{cap:diff-recoils}Differential recoil energy
spectrum for $a=1$ (pure V interaction) and $a=-1$ (pure S
interaction) .}
\end{figure}

A possible admixture of a scalar or tensor type weak interaction
in $\beta$-decay can be studied by determining the $\beta-\nu$
angular correlation. This $\beta-\nu$ angular correlation for
unpolarized nuclei can be written as~\cite{jackson57b}:
\begin{equation}
\omega(\theta_{\beta\nu})\backsimeq1+a\cdot\frac{v_{\beta}}{c}\cos\,\theta_{\beta\nu}\left[1-\frac{\Gamma m}{E}b\right]
\label{eq:beta-nu-angl}
\end{equation}
\noindent where $\theta_{\beta\nu}$ is the angle between the
$\beta$ particle and the neutrino, \emph{E}, $v_{\beta}/c$ and
\emph{m} are the total energy, the velocity relative to the speed
of light and the rest mass of the $\beta$ particle,
$\Gamma=\sqrt{1-(\alpha Z)^{2}}$ with $\alpha$ the fine-structure
constant and $Z$ the nuclear charge of the daughter nucleus,
\emph{b} is the Fierz interference term \emph{}which has
experimentally been shown to be small (e.g.
$\left|b_{F}\right|<0.0044$ at 90\%~C.L.~\cite{hardy05}) and can
as a first approximation be assumed to be zero, and \emph{a} is
the $\beta-\nu$ angular correlation coefficient. Since the S- and
V-interactions lead to Fermi transitions (F) and the A- and
T-interactions to Gamow-Teller transitions (GT), the $\beta-\nu$
angular correlation coefficient \emph{a} can be approximated as
(assuming maximal parity violation for the V- and A-interactions
and real couplings)
\begin{eqnarray}
a_{F} & \simeq & 1-\frac{\left|C_{S}\right|^{2}+\left|C_{S}^{\prime}\right|^{2}}{\left|C_{V}\right|^{2}}\,,\qquad a_{GT}\simeq-\frac{1}{3}\left[1-\frac{\left|C_{T}\right|^{2}+\left|C_{T}^{\prime}\right|^{2}}{\left|C_{A}\right|^{2}}\right]
\label{eq:aF-aGT}
\end{eqnarray}
In the Standard Model, i.e. in the absence of S- and T-type
interactions, $a_{F}=1$ and $a_{GT}=-1/3$. Any admixture of S to V
(T to A) interaction in such a pure Fermi (Gamow-Teller) decay
would result in $a<1$ ($a>-1/3$). A measurement of \emph{a}
therefore yields information about the interactions involved.
However, the neutrino cannot be detected directly and the
$\beta-\nu$ angular correlation thus has to be inferred from other
observables. From the properties of the general Hamiltonian of the
weak interaction (Eq.(\ref{eq:H-general}) and~\cite{wu66}) and of
the Dirac $\gamma$-matrices it can be shown that the two leptons
in $\beta$-decay will be emitted preferably into the same
direction for a \emph{}V~(T) interaction and into opposite
directions for an S~(A) interaction. This will lead to a
relatively large energy of the recoil ion for a V~(T) interaction
and a relatively small recoil energy for an S~(A) interaction
(Fig.~\ref{cap:diff-recoils}). The WITCH experiment aims to
measure the shape of the recoil energy spectrum with high
precision to determine the $\beta-\nu$ angular correlation
parameter \emph{a} and from this deduce a limit on a possible
scalar or tensor admixture in weak interactions.

Due to its feature of being able to measure the recoil ion energy
spectrum in nuclear $\beta$-decay, the WITCH experiment also
provides interesting possibilities for other observables, e.g. one
can also determine F/GT mixing ratios, $Q$-values, $EC/\beta^{+}$
branching ratios and charge state distributions
\cite{kozlov03},~\cite{phd-kozlov}. In addition, the ion cloud in
the decay Penning trap can also be used for $\beta$ and $\gamma$
spectroscopy, this time not using the recoil spectrometer but with
$\beta$ and $\gamma$ detectors added to the WITCH set-up. For
example, a $\beta$ detector on the axis of the decay trap at some
distance behind the trap opens the possibility for $\beta$
spectroscopy with a pure sample without any scattering of the
$\beta$ particles in the source. Adding $\gamma$ detectors around
the center electrode of the trap will make also $\gamma$
spectroscopy possible.

\section{Experiment}

\subsection{Principle}

An experiment to measure the recoil energy spectrum in nuclear
$\beta$-decay faces two major difficulties: (1)~the
$\beta$-emitter is usually embedded in matter, which causes a
distortion of the recoil ion spectrum due to energy losses caused
by ion scattering in the source, and (2)~the recoil ions have very
low kinetic energy rendering a precise energy measurement
difficult.

In order to avoid the first problem and to be as independent as
possible from the properties of the isotopes to be used, the WITCH
experiment uses a double Penning trap~\cite{brown86} structure to
store radioactive ions. The ion cloud in the second trap, the
decay trap, constitutes the source for the experiment, where the
ions are kept for several half-lives, i.e. of the order of 1 to
10\emph{~}s for the cases of interest~\cite{beck03a}.

To solve the second problem and measure the recoil energy spectrum
a retardation spectrometer is used. The working principle of such
a device is similar to the $\beta$-spectrometers used for the
determination of the neutrino rest-mass in Mainz~\cite{picard92}
and Troitsk~\cite{lobashev85}. The WITCH spectrometer consists of
two magnets: the first one providing a field $B_{max}=9$~T, the
second one providing $B_{min}=0.1$~T, and an electrostatic
retardation system. Recoil ions are created in the strong magnetic
field region (i.e. in the Penning trap) and pass on their way to
the detector the region with low magnetic field (i.e. the
retardation section of the spectrometer). Provided that the fields
change sufficiently slow along the path of the ions, their motion
can be considered as adiabatic. According to the principle of
adiabatic invariance of the magnetic flux~\cite{jackson75}
$p_{\bot}^{2}/B=const$, where $p_{\bot}$ is the momentum
projection perpendicular to the magnetic field $B$. From this it
follows that the radial kinetic energies of an ion in the trap
($E_{kin,\,\bot}^{trap}$) and in the retardation section
($E_{kin,\,\bot}^{retard}$) are related as
$E_{kin,\,\bot}^{retard}/E_{kin,\,\bot}^{trap}=B_{min}/B_{max}$.
Thus a fraction
\begin{equation}
1-B_{min}/B_{max}\approx98.9\%
\label{eq:spect-convertion}
\end{equation}
\noindent of the energy of the ion motion perpendicular to the magnetic
field lines will be converted into energy of the ion motion along
the magnetic field lines. The total kinetic energy of the recoil ions
can be probed in the homogeneous region of low magnetic field $B_{min}$
by retarding them with a well-defined electrostatic potential. By
counting how many ions pass the analysis plane for different retardation
voltages, the cumulative recoil energy spectrum can be measured~\cite{beck03a,kozlov03}.

\subsection{Overview of the set-up\label{sub:overview-witch}}

\begin{figure}[H]
\centering

\includegraphics[%
  width=0.99\textwidth,
  keepaspectratio]{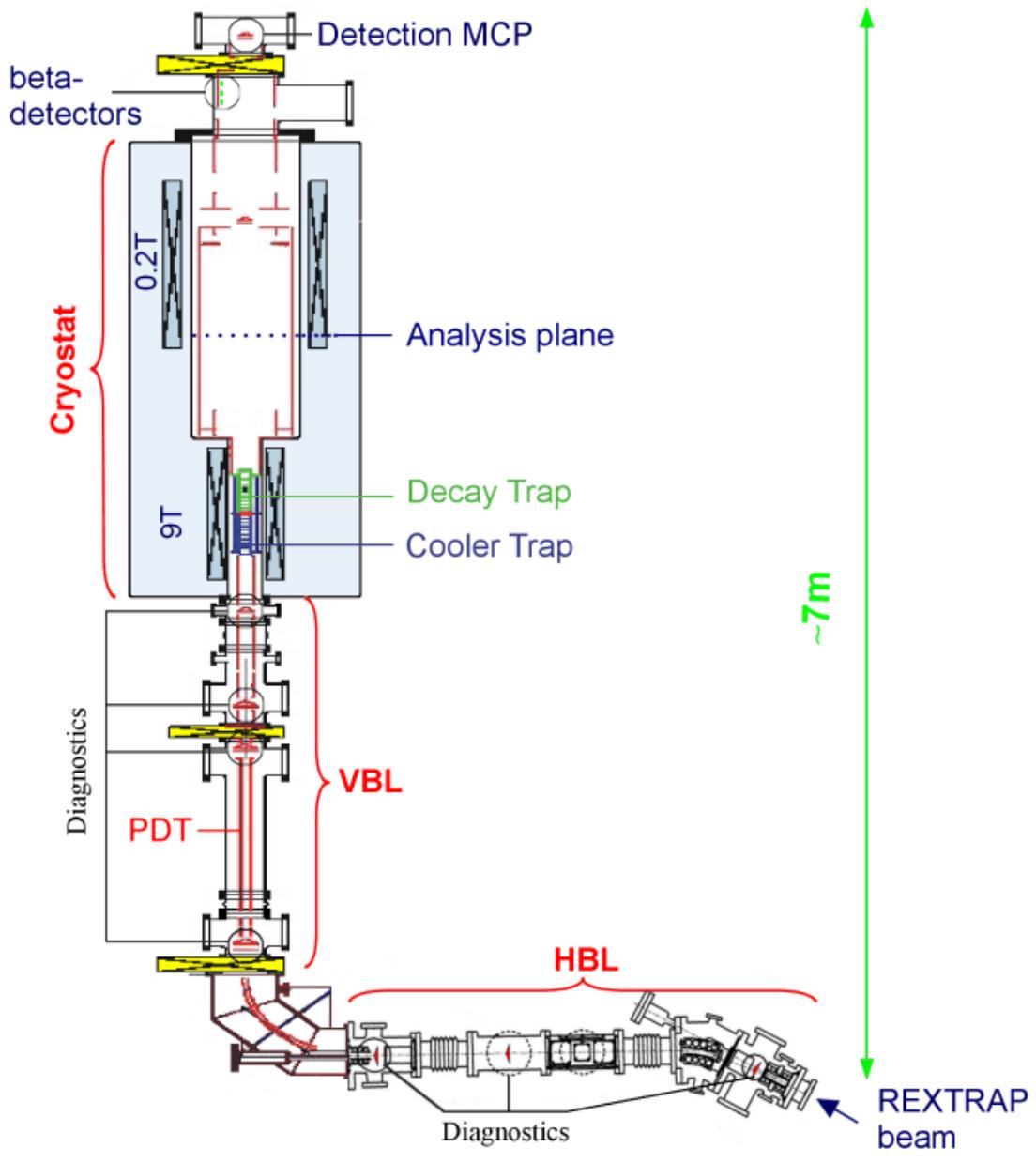}

\caption[General schematic view of the WITCH
set-up]{\label{cap:witch-setup}Schematic view of the WITCH set-up.
The abbreviations used are: HBL horizontal beamline, VBL vertical
beamline, PDT pulsed drift tube. The  HBL \emph{}is $90^{\circ}$
rotated (i.e. the top view of the HBL \emph{}is shown). }
\end{figure}

\begin{figure}[H]
\centering\includegraphics[%
  width=1.0\textwidth,
  keepaspectratio]{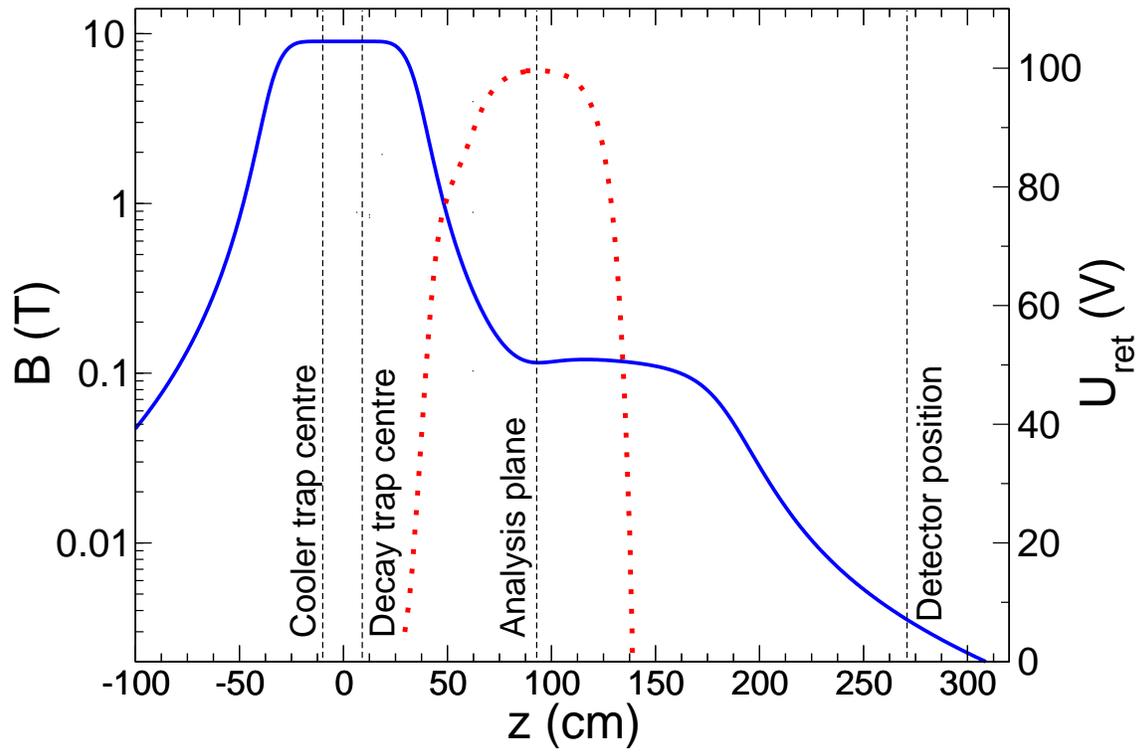}

\caption[Magnetic and electric fields on the axis of the
spectrometer]{\label{cap:Bz-field}Magnetic field (solid line) and
retardation electric field (dashed line) profile on the axis of
the spectrometer (z-axis). The retardation field has been
calculated for $U_{ret}=100$~V. z=0 corresponds to the center of
the 9~T magnet. The positions of the traps, the analysis plane and
the recoil ion detector are also indicated.}
\end{figure}

The general scheme of the set-up can be seen in
Fig.~\ref{cap:witch-setup}. The installation at the ISOLDE
facility at CERN was recently completed. In a first step the ions
produced by ISOLDE \cite{kugler92,kugler00,isolde05a} get trapped
and cooled by REXTRAP~\cite{ames05}. As soon as a sufficient
amount of ions (viz. $10^{6}$ to $10^{7}$ ions) has been collected
by REXTRAP they are ejected as a 60~keV (optionally 30\emph{~}keV)
bunch and are transmitted through the horizontal beamline (HBL) of
WITCH into the vertical beamline (VBL). There the ions are
electrostatically decelerated from 60\emph{~}keV to $\sim$80~eV in
several steps. In order to avoid a high voltage platform a pulsed
cavity is used~\cite{herfurth01b} (named Pulsed Drift Tube, PDT,
in the case of WITCH \cite{delaure05}). When the ion bunch is
inside the PDT the potential of the cavity is switched down over
the range of 60~kV (30~kV) from 52~kV (26~kV) down to -8\emph{~}kV
(-4\emph{~}kV). In this way the kinetic energy of the ions is not
changed while the potential energy is shifted to -8~kV
(-4\emph{~}kV), so that the total energy becomes nearly zero
($\sim$80~eV in practice). The ions can then be captured in the
cooler trap (the first Penning trap of the WITCH set-up), which is
at ground potential. In this cooler trap the ion cloud is prepared
(i.e. cooled and centered) before being ejected through the
pumping diaphragm (which separates the vacua of the two traps)
into the second Penning trap, the decay trap. The latter is placed
at the entrance of the retardation spectrometer. After
$\beta$-decay the recoil ions emitted into the direction of the
spectrometer spiral from the trap, which is in the strong magnetic
field, into the weak field region. In the homogeneous low-field
region the kinetic energy of the ions is then probed by the
retardation potential (Fig.~\ref{cap:Bz-field}). The ions that
pass this analysis region are re-accelerated to $\sim$10~keV to
get off the magnetic field lines. The re-acceleration also ensures
a constant detection efficiency for all recoil ion energies.
Finally, the ions are focused with an Einzel lens onto the
detection micro-channel plate (MCP) detector.
\label{sub:overview-normal}For normalization purposes several
$\beta$-detectors are installed in the spectrometer section
too~(also to check the $\beta$-simulations, see
Sec.~\ref{sub:g4-beta})~\cite{beck03a,kozlov03}. The recoil
spectrum can be measured setting one retardation step for one trap
load (in this case a normalization, e.g. by counting the
$\beta$-particles, is necessary) or scanning all retardation steps
during the same trap load (in which case a correction for the
time-dependence of the count rate due to the isotope half-life has
to be performed). The latter case also allows to avoid a possible
effect of the MCP degradation on the shape of the recoil spectrum.

\subsection{Response function}

Since the WITCH experiment measures the recoil energy spectrum,
good knowledge of the spectrometer response function is of high
importance. The response function has been investigated both
analytically as well as with numerical calculations of the ion
trajectories through the spectrometer~\cite{beck-d-02}. Two
important issues regarding this response function are discussed
below.

\subsubsection{Influence of the residual gas}

The residual pressure in the spectrometer section (and also in the
decay trap) should be as low as possible. Indeed, in order to
measure precisely the recoil spectrum it is very important to
avoid scattering of the recoil ions. The residual gas can also
cause another problem for the experiment, namely charge exchange
which leads to neutralization of the ions such that they can not
be probed anymore by the retardation principle.

\begin{figure}[H]
\centering\includegraphics[%
  width=1.0\textwidth,
  keepaspectratio]{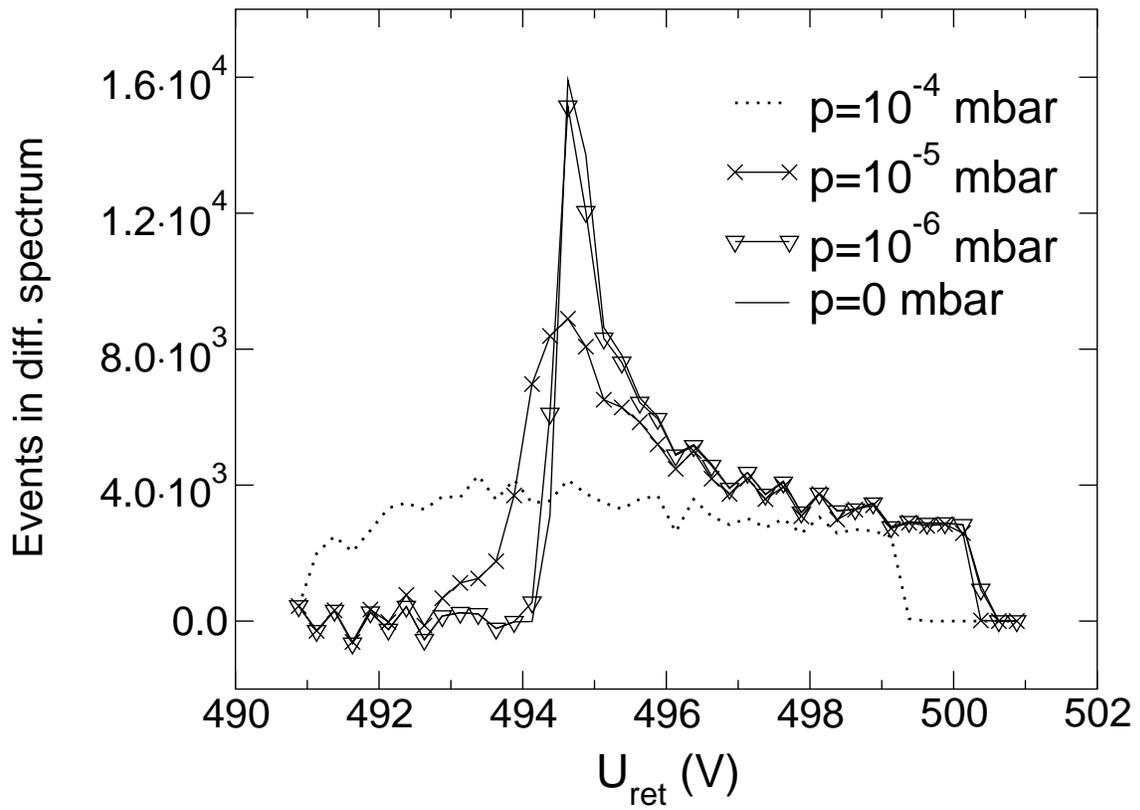}

\caption[Influence of different rest gas pressures on the response
function of the WITCH spectrometer
]{\label{cap:response-rest-gas}Influence of different rest gas
pressures on the response function of the WITCH spectrometer:
0\emph{~}mbar (solid line), $10^{-6}$~mbar (dashed line),
$10^{-5}$~mbar (dash dotted line) and $10^{-4}$~mbar (dotted
line). The simulations are performed for a particle with
$E_{recoil}=500$~eV and mass=50~amu in argon gas.}
\end{figure}

The influence of ion scattering was investigated via simulations
of the ion trajectories, taking into account a Stokes force to
describe the damping of the ion motion. The results are presented
in Fig.~\ref{cap:response-rest-gas}. These simulations were
performed for argon gas in  both traps and in the spectrometer.
The same pressure is used everywhere. For helium gas the indicated
pressures have to be multiplied by a factor of $\approx8$. One can
see that with increasing rest gas pressure the response function
broadens. It can also be noted that for a pressure better then
$10^{-6}$~mbar the energy distribution deviates only a little bit
from the ideal case $p=0$~mbar, meaning that for a successful
recoil spectrum measurement the residual gas pressure has to be
$\le10^{-6}$~mbar in the decay trap and in the spectrometer. Up to
now no estimates were made for the effect of charge exchange on
the WITCH response function.

\subsubsection{Doppler broadening}

\begin{figure}[H]
\centering\includegraphics[%
  width=1.0\textwidth,
  keepaspectratio]{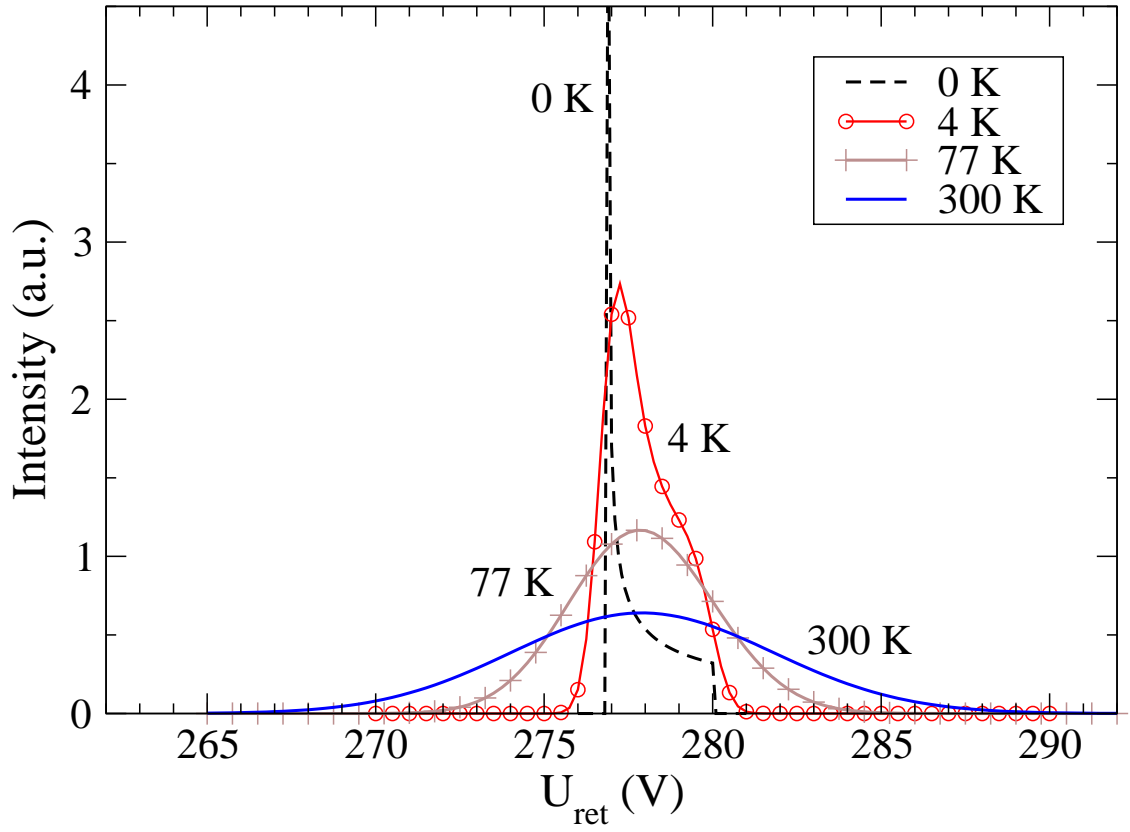}

\caption[Doppler broadening of the response function of the WITCH
spectrometer]{\label{cap:response-doppler}Doppler broadening of
the response function of the WITCH spectrometer for different
temperatures of the ion cloud. Calculations performed for
$E_{recoil}=280$~eV without taking into account the rest gas
pressure.}
\end{figure}

The ideal response function was derived assuming that the ions are
at rest. The real situation, however, differs since the ion cloud
in the decay trap is at least at room temperature. The velocity of
the recoil ion is thus the superposition of the velocity of the
mother nucleus in the trap and the velocity obtained due to the
beta decay. This definitely affects the recoil spectrum. Assuming
a Maxwellian ion velocity distribution and taking into account
that the energy at room temperature (0.025\emph{~}eV) is much
smaller than the recoil energy \emph{O}(100\emph{~}eV), it can be
shown \cite{phd-delaure} that the energy shift $\epsilon$ due to
Doppler broadening results in the Gaussian distribution:

\begin{equation}
f_{\epsilon}=\frac{1}{\sqrt{2\pi}\sigma_{\epsilon}}\exp\left(-\frac{\epsilon^{2}}{2\sigma_{\epsilon}^{2}}\right),\label{eq:doppler-shift}\end{equation}

\noindent with

\begin{equation}
\sigma_{\epsilon}=\sqrt{2E_{kin}^{recoil}kT}\label{eq:doppler-sigma}\end{equation}

\noindent where $E_{kin}^{recoil}$ is the recoil ion energy
obtained in the beta decay, \emph{k} is the Boltzmann constant and
\emph{T} is the ion cloud temperature. Eq.(\ref{eq:doppler-sigma})
yields, for e.g. $E_{kin}^{recoil}=280$~eV and $T=300$~K,
\emph{}$\sigma_{\epsilon}=3.8$~eV. This means that the
mono-energetic response function broadens significantly if the ion
cloud is at room temperature. To get the real response function
for the WITCH spectrometer one has to fold the ideal
mono-energetic response function with this broadened energy
distribution. Fig.~\ref{cap:response-doppler} shows the WITCH
response function thus obtained for different ion cloud
temperatures. As one can see, down to liquid nitrogen temperature
(i.e. 77~K) the shape of the response function is defined by
Doppler broadening and only at the temperature of liquid helium
(i.e. 4~K) it approaches the shape of the ideal response function.
It would thus be interesting to consider to cool (at least to
77\emph{~}K) both Penning traps.

\begin{figure}[H]
\centering\includegraphics[%
  width=1.0\textwidth,
  keepaspectratio]{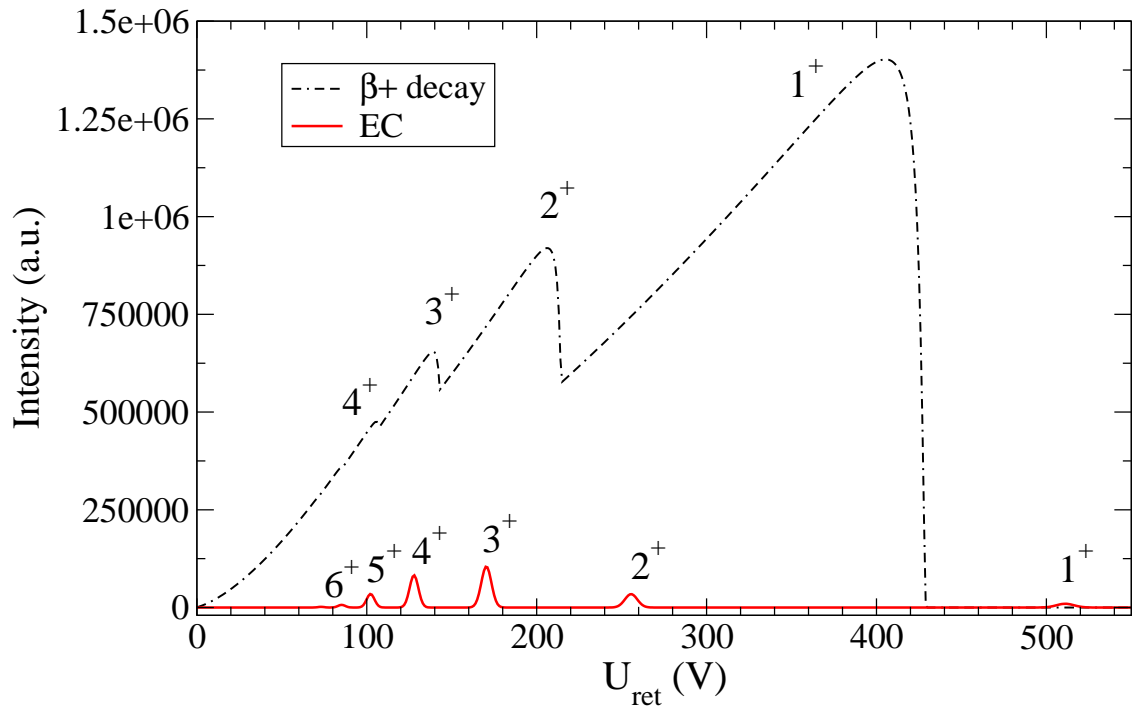}

\caption[Differential recoil spectrum calculated for the
$\beta^{+}$-decay of $^{38m}$K]{Differential recoil spectrum
calculated for the $\beta^{+}$-decay of $^{38\textnormal{m}}$K.
The charge distribution for the $\beta^{+}$-decay is taken from
\cite{gorelov00}, while the charge state distribution for electron
capture is unknown for this isotope. In order to still get a
qualitative idea the known charge state distribution for $^{37}$Ar
is therefore shown (from \cite{snell55})}

\label{cap:spectrum-witch}
\end{figure}

The response function can be determined experimentally by
measuring recoil ions from electron capture decay (EC), which
leads to a mono-energetic peak that is, if it accompanies
$\beta^+$ decay, above the endpoint energy of the continuous
recoil spectrum from $\beta^{+}$ decay
(Fig.~\ref{cap:spectrum-witch}, see Sec.~\ref{sub:recoil_spec}).
By measuring these EC peaks for several suitable nuclides with
different decay energies, an energy calibration of the
spectrometer can be performed as well.

\subsection{Recoil spectrum\label{sub:recoil_spec}}

Electron shake-off after the $\beta$-decay~\cite{snell68} will
cause the daughter ions to have different charge states $q=n\cdot e$.
The recoil ions with energy $E_{kin}^{recoil}$ and charge \emph{q}
will appear in the measured spectrum at a retardation voltage $U_{ret}=E_{kin}^{recoil}/q$
due to the retardation principle. The measured spectrum will thus
be a superposition of the spectra of the various charge states, each
with different endpoints $U_{0n}=E_{0}^{endpoint}/(n\cdot e)$, $n\ge1$,
where $E_{0}^{endpoint}$ is the endpoint energy of the recoil energy
spectrum for \emph{n} = 1 (Fig.~\ref{cap:spectrum-witch}). Consequently,
when measuring the full recoil spectrum up to the endpoint energy
$E_{0}^{endpoint}$, the upper half of the spectrum will consist purely
of events from charge state \emph{n} = 1. This upper half of the recoil
spectrum is therefore the most interesting part for analyzing the
spectrum shape~\cite{kozlov03}. Note that the EC peaks of different
charge states \emph{n} for the same nuclide will also appear in the
spectrum at different energies $E_{EC}/n$, $n\ge1$.

It is also planned to check the dependence of the shake-off
probability on the recoil ion energy, as has been seen in $^6$He
$\beta^-$ decay~\cite{carlson63}. This dependence is expected to
be even larger in $\beta^+$ decay~\cite{scielzo03}. This effect to
first order distorts the recoil ion spectrum by $(1+s\cdot
E_{recoil})$. The idea is to use the upper half of the 1$^+$
charge state spectrum to fit simultaneously $s$ and $\beta - \nu$
correlation coefficient $a$. A possible dependence of the
shake-off process on the recoil energy can be checked by fitting
spectra obtained for different charge states to $a$ (not including
the $s$ parameter).

\subsection{Achievable precision\label{sub:precision}}

\begin{figure}[H]
\centering\includegraphics[%
  width=1.0\textwidth,
  keepaspectratio]{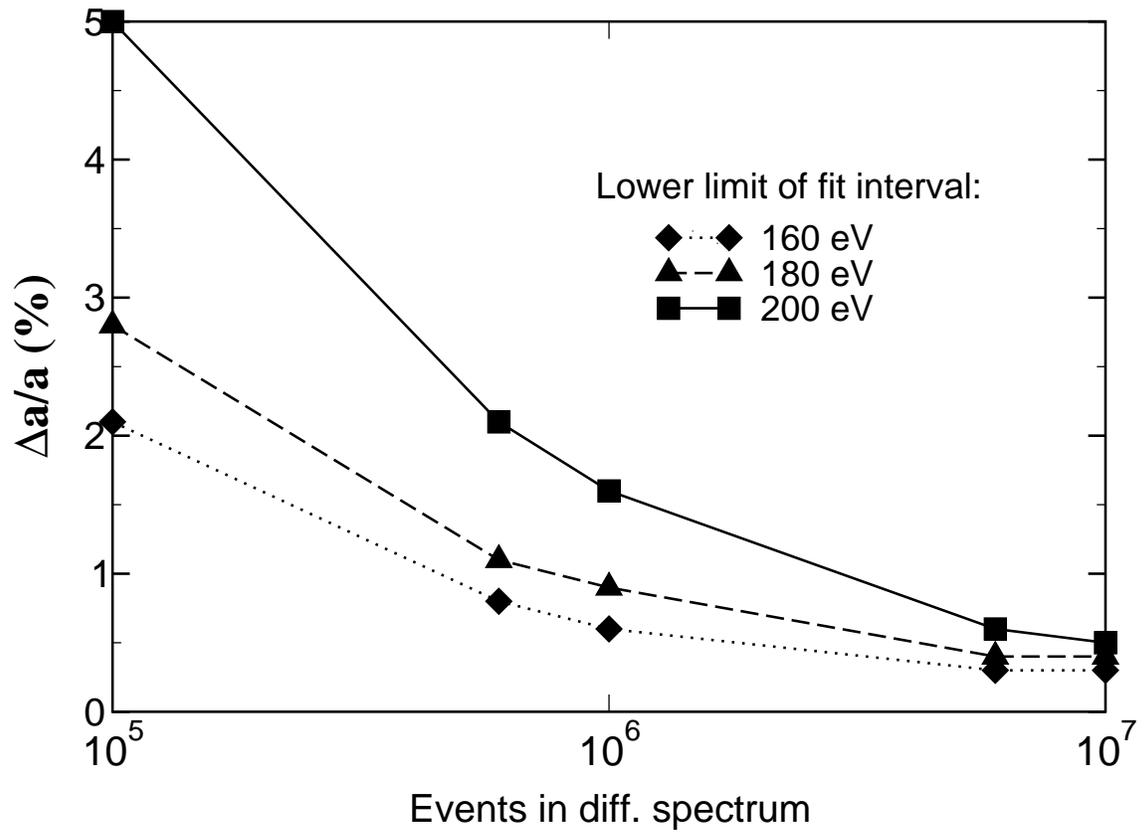}

\caption[Estimated precision on the $\beta-\nu$ angular
correlation coefficient \emph{a} as a function of the total number
of events N in the differential recoil
spectrum]{\label{cap:precision-a}Estimated precision on the
$\beta-\nu$ angular correlation coefficient \emph{a} (for
$^{26\textnormal{m}}$Al, C.L.=68.3\%) as a function of the total
number of events N in the differential recoil spectrum when three
different energy intervals near the endpoint (at 280.6~eV for
$^{26\textnormal{m}}$Al) are considered for analysis.}
\end{figure}

A series of random integral recoil spectra have been generated
from which the $\beta-\nu$ angular correlation parameter \emph{a}
was fitted. The response function of the spectrometer was
approximated by Gaussian with $\sigma=$1\%. No $\beta$ background
 and no dependence of the shake-off on the recoil
ion energy were considered. By changing the fit interval and the
bin width the influence on the achievable precision for \emph{a}
was studied. These simulations (Fig.~\ref{cap:precision-a}) show
that, to reach a precision of $\Delta a=0.005$, the total number
of events in the differential energy spectrum should be
$N=10^{7}\div10^{8}$ and a minimum of $n_{0}=20$ channels (i.e.
retardation steps) in the upper half of the spectrum seems to be
sufficient. Taking $N=10^{8}$, $n_{0}=20$, the number of ions in
one trap load $N_{load}=10^{6}$ and the efficiency parameters for
a fully optimized set-up (Table~\ref{cap:efficiency-witch}), one
can find that the measurement time needed to reach a precision of
$\Delta a=0.005$ (Fig.~\ref{cap:precision-a}) is
$t_{meas}\approx3.6$~days for $^{35}$Ar ($t_{1/2}=1.78$~s) and
$t_{meas}\approx13$~days for $^{26\textnormal{m}}$Al
($t_{1/2}=6.35$~s) \cite{beck03a}. In fact, the results of the
simulations presented in Fig.~\ref{cap:precision-a} show that the
required precision of $\Delta a=0.005$ might be achieved already
with $N=10^{7}$ which means that the measurement time reduces
correspondingly, i.e. $t_{meas}\approx8.6$~hours for $^{35}$Ar and
$t_{meas}\approx1.3$~days for $^{26\textnormal{m}}$Al.

\subsection{GEANT4 $\beta$-particle simulations\label{sub:g4-beta}}

\begin{figure}[H]
\centering\includegraphics[%
  width=0.99\textwidth,
  keepaspectratio]{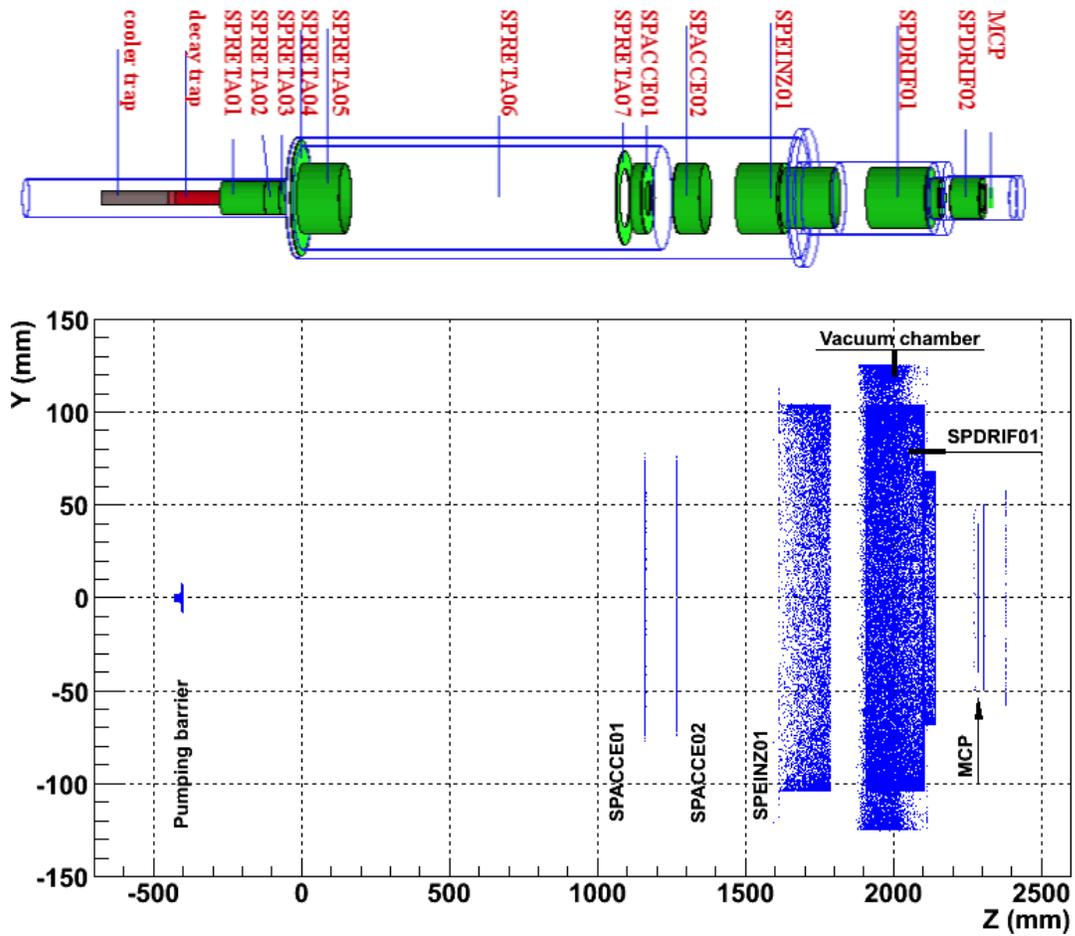}

\caption[2D plot of the points where
\$\textbackslash{}beta\$-particles hit the set-up for the first
time]{Plot of the points where $\beta$-particles hit the set-up
for the first time. $^{35}$Ar spectrum, 5~mm cloud, isotropic
distribution. The center of the decay trap is at -323~mm. On top
 the WITCH set-up in the GEANT4 simulation program is shown.
 The different spectrometer (SP) electrodes are indicated. \label{cap:betas-g4-1st-hit}}
\end{figure}

As was already mentioned in Sec.~\ref{sub:overview-normal}
normalization detectors (to count $\beta$-particles) are available
in the set-up too. Simulations with the GEANT4 simulation
package~\cite{geant4-03,geant4-web} were performed to find the
best suited place for installing these detectors in order to
provide good statistics for the normalization. The scheme of the
simulated set-up is presented in
Fig.~\ref{cap:betas-g4-1st-hit}~(\emph{top}). The double Penning
trap structure is described as a set of two solid copper cylinders
(inner diameter 40\emph{~}mm) separated by a copper pumping
diaphragm with an opening of 4~mm diameter. All spectrometer
electrodes as well the walls of the lower and upper magnet bore
tubes are made from stainless steel. For the MCP detector a very
simplified model is used, i.e. a solid disk of quartz material
($\textnormal{Si}\textnormal{O}_{2}$), 80\emph{~}mm in diameter
and 1\emph{~}mm thick.

The program was run for the case of $^{35}$Ar, with an isotropic
$\beta$-particle emission distribution and a 5\emph{~}mm radius
ion cloud in the center of the decay trap. The positions and
elements of the set-up which are hit first by the
$\beta$-particles after they leave the decay trap were determined.
This is shown in Fig.~\ref{cap:betas-g4-1st-hit}. One can
recognize the shape of certain electrodes and also the pumping
diaphragm (at nearly -400\emph{~}mm). The largest number of hits
is in the pumping barrier: nearly 80\% of the betas going in this
direction hit the barrier. The second most hit element is the
SPDRIF01 electrode. Taking into account that the size of the
pumping barrier is much smaller than the size of the SPDRIF01
electrode, one can conclude that the pumping barrier is the most
suitable place to install the $\beta$-detectors.

Another important result of these simulations is the estimate for
the $\beta$-back\-ground on the recoil ion detector. This MCP
detector is introduced in the program as a very primitive model.
Nevertheless, this allows to check how many particles will arrive
on the detector. It appears that this amount is actually rather
large: from a total of 250~000 simulated events 8608 betas arrive
on the 8\emph{~}cm diameter \emph{}MCP detector. This number has
to be compared to the number of ions arriving on the same
detector. Considering the efficiencies for an ideal set-up
(Table~\ref{cap:efficiency-witch}) it can be estimated that about
10~000 ions will arrive on the MCP detector. However, it might be
that an MCP of 4\emph{~}cm diameter is already sufficient for
detecting all ions. Since the $\beta$-background events equally
cover the surface of the MCP, a two times smaller diameter of the
detector means a reduction of the $\beta$-background by a factor
of 4, corresponding to 2152 beta-particle events. This estimate of
the $\beta$-background is not yet complete however, as one still
has to take into account the different detection efficiency of the
MCP for ions of \emph{O}(10\emph{~}keV) energy (the recoil ions
are accelerated onto the MCP, see Sec.~\ref{sub:overview-witch})
and $\beta$-particles of \emph{O}(MeV) energy. The model used for
the MCP detector is too primitive to answer this question.
Measurements performed by the TRIUMF and Berkeley groups
\cite{behr05,scielzo05} showed that this efficiency is actually
very close to the one for ions (i.e. the absolute MCP detection
efficiency is about 50$\div$75\% for $\beta$-particles of
\emph{O}(MeV) energy, compared to about 60\% for ions). This means
that if one assumes the MCP registration efficiency for betas and
ions to be both 60\%, the MCP detector will register
\emph{$\sim$1290} $\beta-$particles against \emph{6~000} ion
events in the full recoil ion spectrum.

The first estimates~\cite{phd-kozlov} show that in order to
achieve $\Delta a=0.005$ with such $\beta$ background an increase
of the measurement time by about one order of magnitude might be
necessary. However, if one considers that $10^7$ counts in the
differential spectrum (not including $\beta$ background) is
sufficient (Fig.~\ref{cap:precision-a}, Sec.~\ref{sub:precision}),
the required precision on the $\beta-\nu$  correlation ($\Delta
a=0.005$) might still be achieved in a realistic time period.
These simulations, for sure, have to be checked experimentally and
the issue of the MCP detection efficiency for ions and betas still
has to be investigated in more detail too.

\section{Tests performed}

The WITCH set-up was completed only recently and commissioning was
started already during the last phase of the installation. In most
of the tests the off-line REXTRAP ion source (delivering a $^{39}$K
ion beam) was used. The tests performed are discussed in this section.

\subsection{HBL}

The function of the \emph{}HBL \emph{}is to transfer the ion beam
from the REXTRAP set-up into the WITCH vertical beamline as
efficient as possible. Since the ISOLDE facility (and
respectively, REXTRAP) is operated usually at either 60\emph{~}kV
or 30\emph{~}kV, tests were done using both high voltage settings.
The tests performed showed that the tuning can be done in such a
way that there is no significant loss of beam intensity through
the HBL, i.e. nearly 100\% transmission efficiency is obtained. To
tune the voltages and monitor the beam, the three diagnostics in
the HBL and the first one in the \emph{}VBL are used.

\subsection{VBL}

The main purpose of the VBL tests that were performed till now was
to prove the functionality of the PDT, check its efficiency and investigate
and optimize the injection of the ion beam into the WITCH magnetic
field.

\begin{figure}[H]
\centering\includegraphics[%
  width=1.0\textwidth,
  keepaspectratio]{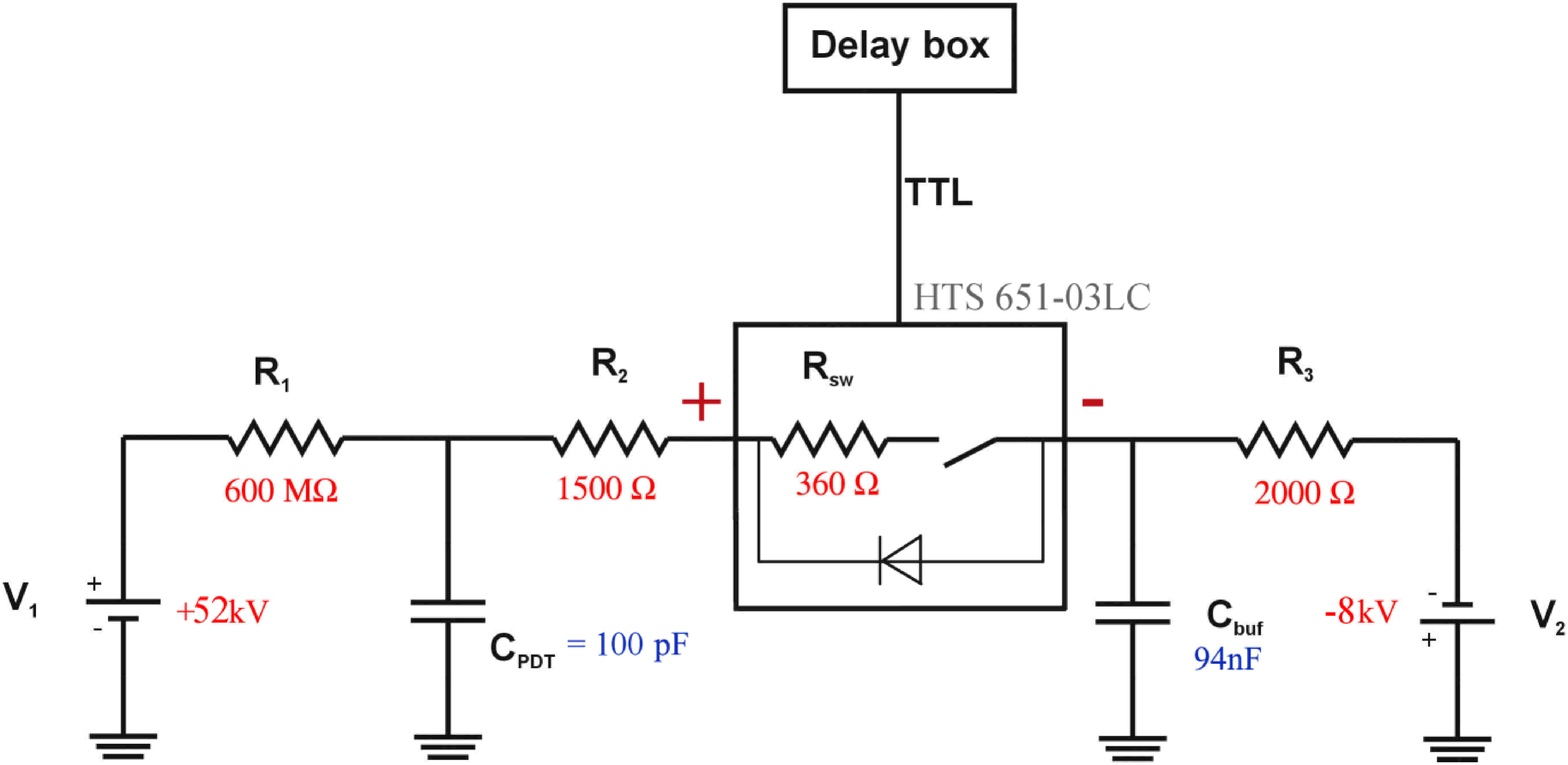}

\caption{\label{cap:hv-switch-scheme}Electrical scheme of the HV
switch system for 60\emph{~}kV\emph{.}}
\end{figure}

\begin{figure}[H]
\centering\includegraphics[%
  width=1.0\textwidth,
  keepaspectratio]{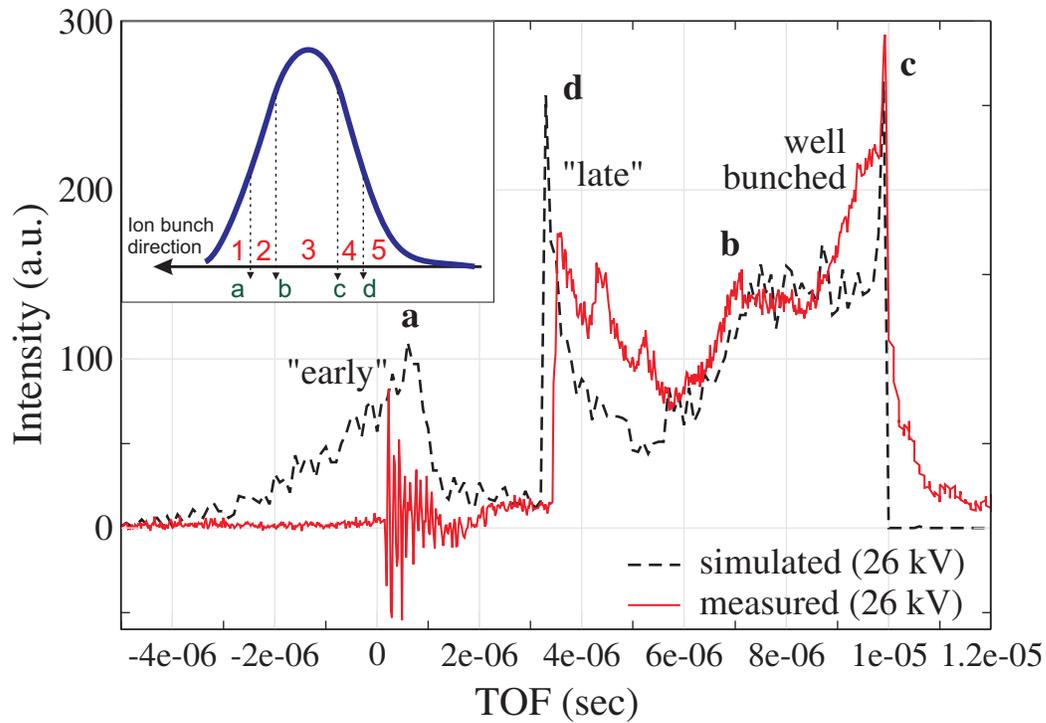}

\caption[TOF spectra of $^{39}$K after the PDT (third VBL MCP
detector)]{\emph{\label{cap:pdt-tof}}TOF spectra of $^{39}$K after
the PDT. \emph{}The simulation is for a HV switch time constant =
0.2~$\mu s$. The measured spectrum is the MCP signal inverted and
scaled to the simulated spectrum. Measurement and simulation are
both for HV PDT = 26\emph{~}kV. The zero of the TOF\emph{-axis}
corresponds to the start of the HV switching. In top left corner
the schematic of the original ion bunch is shown: 1) "early" ions,
2) partially bunched "early", 3) well bunched ions, 4) partially
bunched "late", 5) "late" ions. a, b, c and d correspond to the
transitions between different parts.}
\end{figure}

From the values for the resistances and capacitances in the HV
switch box scheme (Fig.~\ref{cap:hv-switch-scheme}) the HV
switching time constant $\tau_{PDT}$ was estimated to be 0.18$\mu
s$. This provides a pulse down time $t_{switch}$ of the PDT of
about $1.3\,\mu s\;(1.2\,\mu s)$ for a 60\emph{~}kV (30\emph{~}kV)
ISOLDE beam, with $t_{switch}$ being defined such that the energy
of the ions after the switching differs less than 50~eV from the
required value~\cite{delaure05}. The travel time of 30\emph{~}keV
\emph{}$^{39}$K$^+$ ions in the PDT (for a combination of HV =
\mbox{+26/-4~kV}) is $\sim5\,\mu s$. This gives an $\sim3.7\,\mu
s$ time window to pulse down the ion beam. However, the ion bunch
ejected from the REXTRAP set-up has a time structure that is
longer than this time window (the typical bunch length is
$\sim10\,\mu s$), meaning that it is not possible to pulse the
complete bunch but at best only about 43\% of it for a combination
of HV = +26/-4\emph{~}kV~\cite{delaure05}. The original bunch can
thus be considered to consist of different parts: 1) ions which
will pass through the PDT \emph{}before the pulsing down starts,
2) an intermediate part containing partially bunched ions, 3) well
bunched ions, 4) \emph{}another intermediate part of partially
pulsed down ions and, finally, \emph{}5) ions which enter the PDT
when this is already at low voltage (Fig.~\ref{cap:pdt-tof},
\emph{top left corner}). For the cases 1) and \emph{}5) these ions
will have $\sim60$~keV (30~keV) after the PDT, i.e. they are much
faster than the well-bunched ions. This leads to the time
structure of the signal after the PDT shown also in
Fig.~\ref{cap:pdt-tof}. In this figure the simulated (using the
SIMION simulation routine~\cite{dahl95}) and measured spectra are
compared. As can be seen, good agreement between simulations and
the measurement is obtained.

The efficiency \emph{}of the \emph{}PDT, $\eta_{PDT}$, was
measured using two diagnostic MCP detectors: one in front of the
PDT and one behind it at $\sim$40\emph{~}cm from the exit of the
PDT. The resulting efficiency $\eta_{PDT}=8\%$ is less than the
expected value of $\sim$43\% but is of the right order of
magnitude~\cite{delaure05}.

From the signal of the MCP detector behind the trap structure
(i.e. in the magnetic field) the overall efficiency of the
vertical beamline, including the efficiency of the pulsed drift
tube, was found to be between 0.1\% and 1\% (for these
measurements ions were not trapped but only sent through the
traps)~(Table~\ref{cap:efficiency-witch}). Additional measurements
were performed in the mean time with improved MCP diagnostic
detectors~\cite{coeck05b}, i.e. with several Ni meshes in front to
reduce the incoming beam intensity, and a grid-anode to determine
the beam size. These have revealed a significant decrease of beam
quality after the PDT. This is being looked into in more detail
now.

\subsection{Traps}

In the initial phase, i.e. for a simple mode WITCH operation and tests
of the spectrometer, a sophisticated use of the traps is not required.
It was therefore not our aim to perform already now a detailed investigation
of their properties. Rather we wanted to verify the simple operation
of the traps, i.e. try out basic trap mechanisms, understand the behavior
of ions in the traps and, if possible, optimize the parameters in
order to find a suitable trapping regime.

\subsubsection{Buffer gas cooling}

An important step in verifying the operation of Penning traps is to
check the buffer gas cooling procedure since this is usually a first
main step in the ion cooling. The buffer gas used is high purity $^{4}$He
(quality: Helium 57 or >99.9997\%), but the transfer line is not yet
equipped with a cold trap or other purification system. The buffer
gas pressure measurement is done at a point after the gas dosing valve
(Pfeiffer Vacuum, RME005) and before the transfer line enters into
the vacuum chamber. With this installation the buffer gas pressure
is regulated via a feedback loop.

\begin{figure}[H]
\begin{minipage}[c]{1.0\columnwidth}%
\centering\includegraphics[%
  height=0.45\textwidth,
  keepaspectratio]{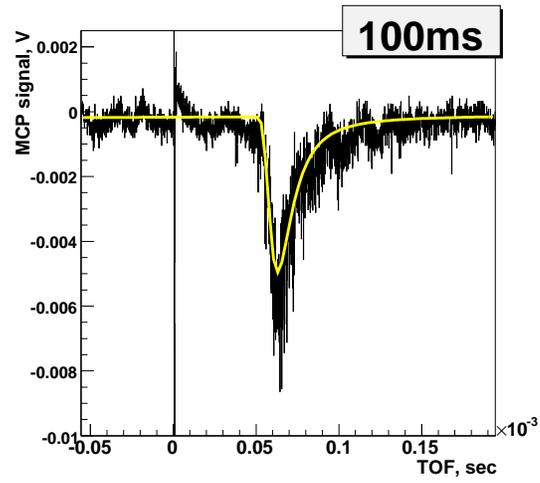}\end{minipage}%

\begin{minipage}[c]{1.0\columnwidth}%
\vspace{10pt}

\centering\includegraphics[%
  height=0.45\textwidth,
  keepaspectratio]{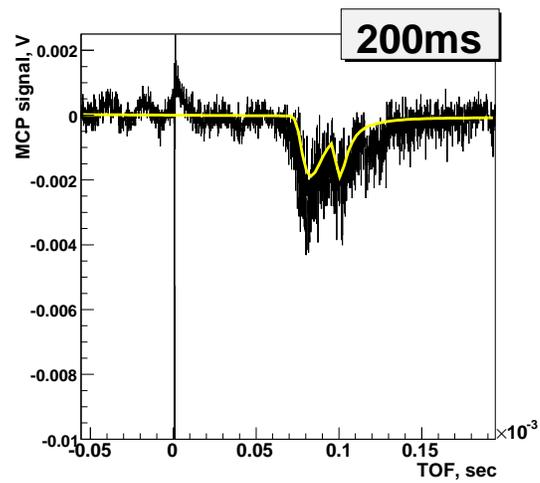}\end{minipage}%

\begin{minipage}[c]{1.0\columnwidth}%
\vspace{10pt}

\centering\includegraphics[%
  height=0.45\textwidth,
  keepaspectratio]{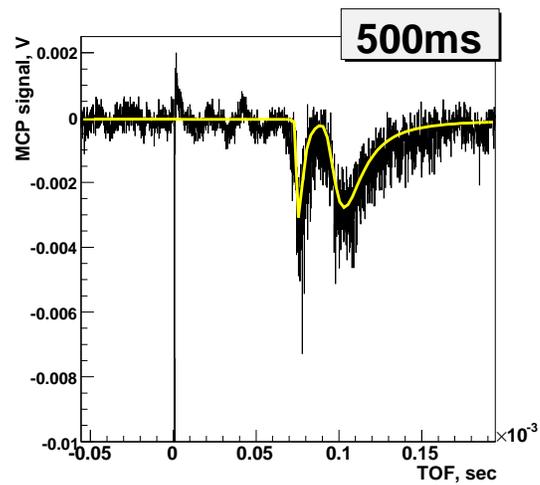}\end{minipage}%

\caption[Oscilloscope TOF pictures observed with an MCP for
different cooling times]{\label{cap:osc-signal-5mbar}Oscilloscope
TOF pictures observed with an MCP for different cooling times
(100\emph{~}ms, 200\emph{~}ms, 500\emph{~}ms) in the cooler trap.
The buffer gas pressure was 5\emph{~}mbar (at the gas dosing valve
position).}
\end{figure}

A scan of the MCP signal as a function of the cooling time was
performed for a buffer gas pressure of 5\emph{~}mbar at the gas
dosing valve position. $^{39}$K$^+$ ions are trapped in the cooler
trap, cooled there for some time and then sent (without capturing)
through the decay trap onto the MCP detector behind the WITCH trap
structure. It was found that after $\sim$200\emph{~}ms \emph{}of
cooling the MCP signal splits in two different peaks
(Fig.~\ref{cap:osc-signal-5mbar}). This means that other ions then
$^{39}$K$^+$ are present too, for instance because the buffer gas
is not clean enough. Using dipole excitation of the reduced
cyclotron motion ($\nu_{+}$), the first peak was identified as
mass 19 (H$_{2}$OH$^{+}$) while the second one corresponds to
$^{39}$K$^+$. Qualitatively the effect of the cooling can be seen
in Fig.~\ref{cap:osc-signal-5mbar}: after 100~ms of cooling the
TOF position of the $^{39}$K$^+$ peak is $\sim70\,\mu s$ while
after 200~ms of cooling the peak appears at $\sim100\,\mu s$.

\subsubsection{Excitations}

\begin{figure}[H]
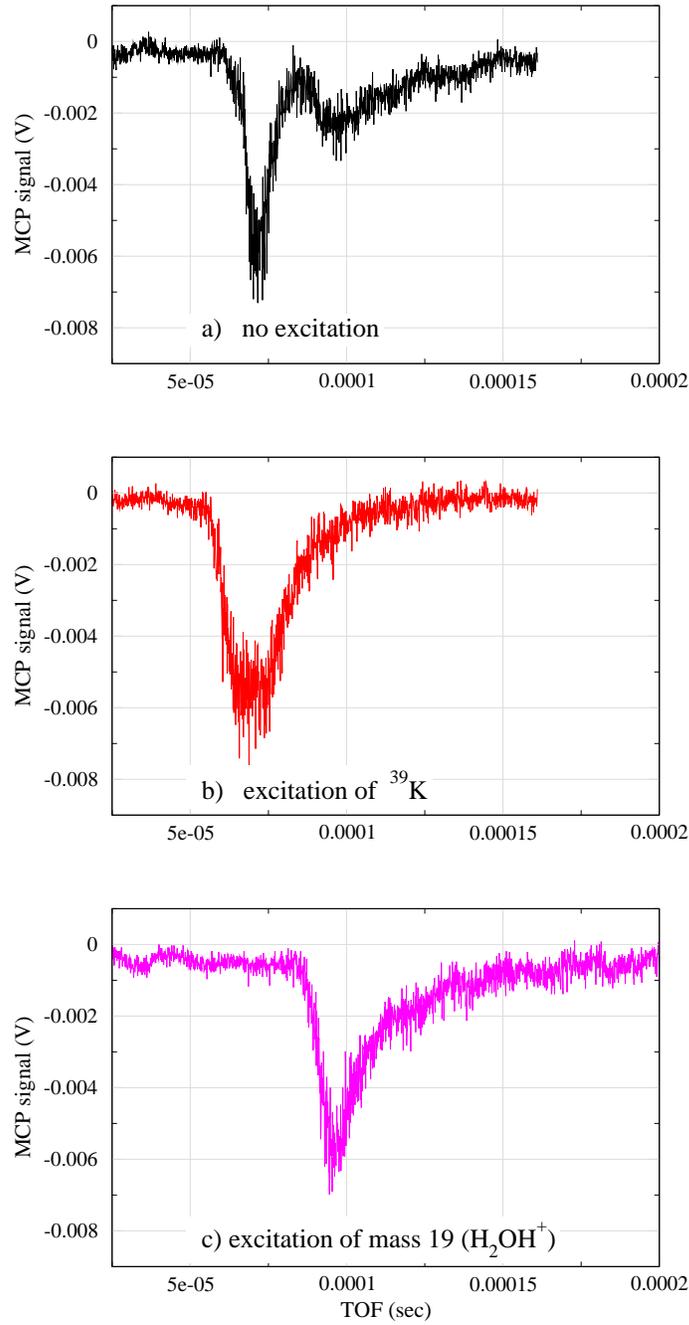

\begin{minipage}[c]{1.0\columnwidth}%
\centering\includegraphics[%
  height=0.37\textwidth,
  keepaspectratio]{graphs/ion-ident-no_exc.eps}

~\end{minipage}%

\begin{minipage}[c]{1.0\columnwidth}%
\centering\includegraphics[%
  height=0.37\textwidth,
  keepaspectratio]{graphs/ion-ident-39K_exc.eps}

~\end{minipage}%

\begin{minipage}[c]{1.0\columnwidth}%
\centering\includegraphics[%
  height=0.37\textwidth,
  keepaspectratio]{graphs/ion-ident-19_exc.eps}

~\end{minipage}%

\caption[Identification of different peaks based on the dipole
excitation at the reduced cyclotron
frequency]{\label{cap:peak-indent}Identification of different
peaks from the TOF observed with an MCP detector after different
excitations in the cooler trap. Ions are a) first cooled for
200\emph{~}ms, without exciting them and then b) dipole excited at
$\nu_{rf}=\nu_{+}(^{39}\textnormal{K})=2367100\textnormal{\, Hz}$
($A_{\nu_{+}}=2$~V) during 100\emph{~}ms and c) excited at the
reduced cyclotron frequency $\nu_{rf}=4845000\textnormal{\,
Hz}\sim\nu_{+}$(mass~19) ($A_{\nu_{+}}=0.5$~V) during
100\emph{~}ms (in this last case the ions were first cooled during
400\emph{~}ms). The peak corresponding to the excited mass
disappears. The magnetic field was 6~T. The He buffer-gas pressure
was 5\emph{~}mbar. The increase of the MCP signal after removing
the first peak (case (c)) could be related to MCP effects (see
Sec.~\ref{sub:MCP-regime})}
\end{figure}

In Fig.~\ref{cap:peak-indent} the oscilloscope pictures
corresponding to the dipole $\nu_{+}$ excitation are shown. With
no applied excitation two peaks (H$_{2}$OH$^{+}$ and $^{39}$K$^+$)
are visible. Next, the following scheme is used: first the ions
are cooled by collisions with the buffer gas atoms in the cooler
trap during 200\emph{~}ms, thereafter they are excited at
$\nu_{+}(^{39}\textnormal{K})=2367100$~Hz (in a 6\emph{~}T
B-field) for 100\emph{~}ms with an amplitude of $A_{\nu_{+}}=2$~V
and, finally, they are extracted
(Fig.~\ref{cap:peak-indent}\emph{b}). As can be seen the
$^{39}$K$^+$ peak disappears, i.e. with
$\nu_{+}(^{39}\textnormal{K})$ excitation the $^{39}$K$^+$ ions
were brought to a radius larger than the radius of the pumping
diaphragm. In another test the ions in the first peak were excited
at the reduced cyclotron frequency $\nu_{rf}=4845000\textnormal{\,
Hz}\simeq\nu_{+}$(mass~19) ($A_{\nu_{+}}=0.5$~V) during
100\emph{~}ms (in this case the ions were first cooled during
400\emph{~}ms) (Fig.~\ref{cap:peak-indent}\emph{c}). The same
effect is obtained, i.e. the corresponding peak disappears. The
increase of the MCP signal after removing the first peak could be
related to MCP effects (see Sec.~\ref{sub:MCP-regime}). No
systematic scan of the number of ions ejected from the trap as a
function of the excitation frequency was performed as yet, but
only a qualitative study of the TOF oscilloscope spectrum. These
tests nevertheless show that the $\nu_{+}$ excitation works for
WITCH and allows to separate different masses.

\begin{figure}[H]
\begin{minipage}[c]{1.0\columnwidth}%
\centering\includegraphics[%
  width=1.0\textwidth,
  keepaspectratio]{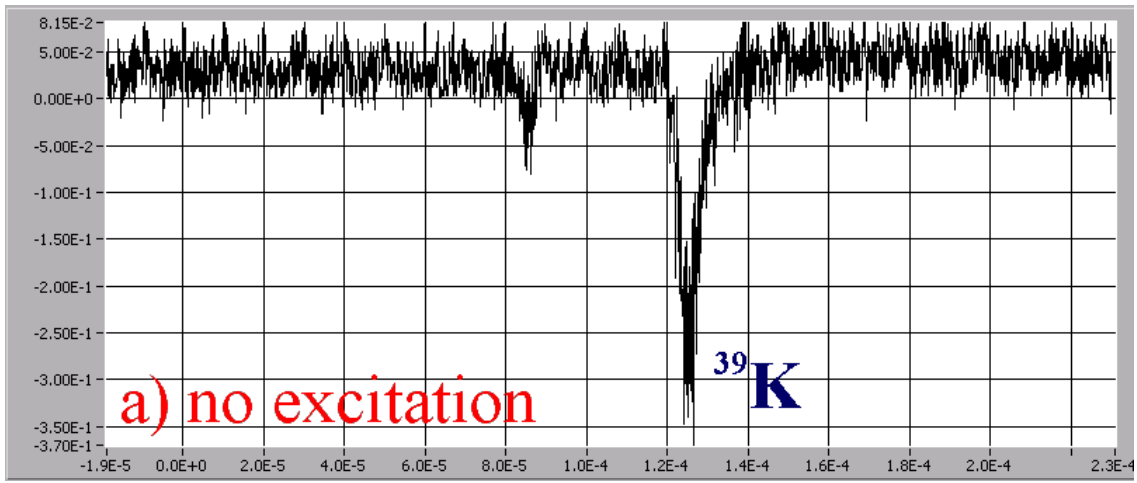}

~\end{minipage}%

\begin{minipage}[c]{1.0\columnwidth}%
\centering\includegraphics[%
  width=1.0\textwidth,
  keepaspectratio]{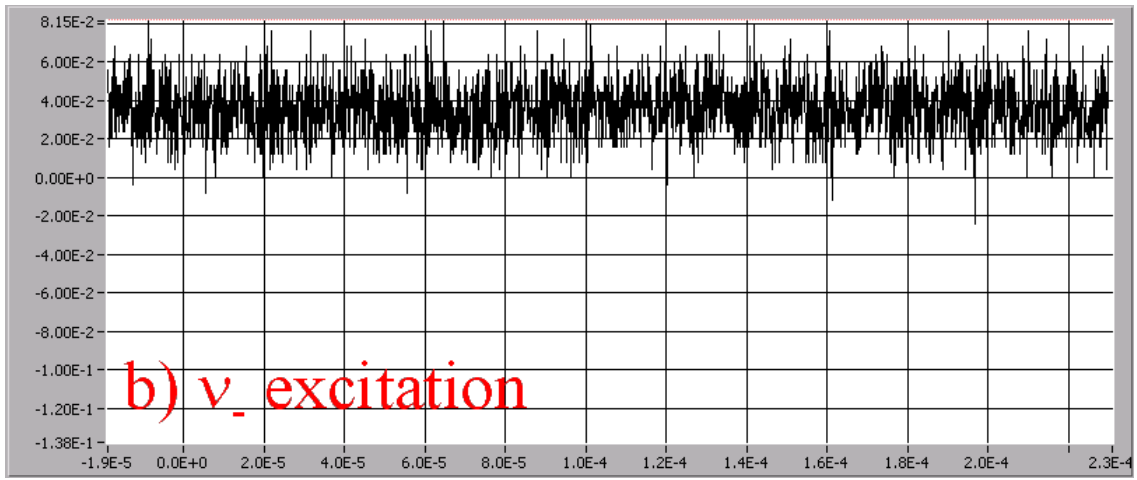}

~\end{minipage}%

\begin{minipage}[c]{1.0\columnwidth}%
\centering\includegraphics[%
  width=1.0\textwidth,
  keepaspectratio]{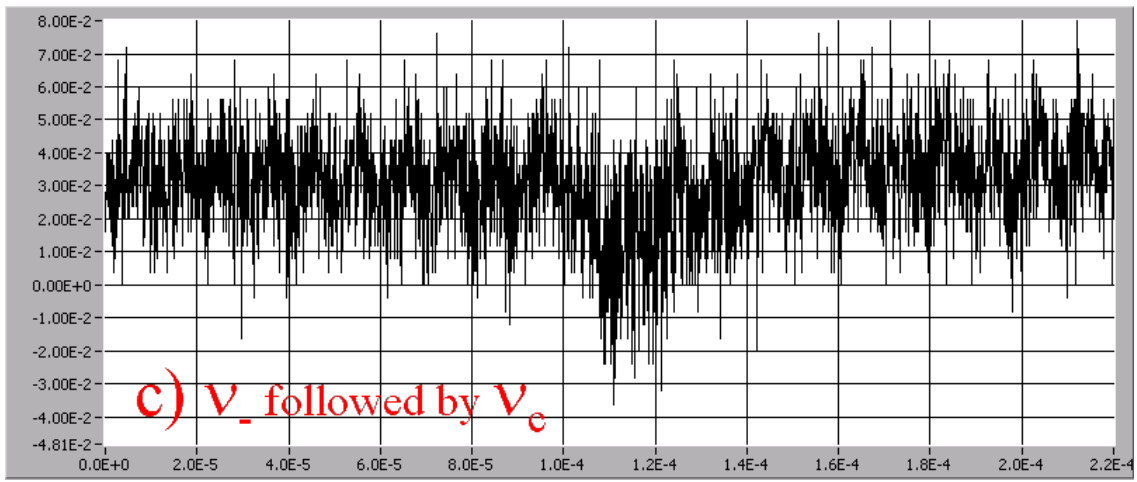}

~\end{minipage}%

\caption[Mass selective cooling of
$^{39}$K]{\label{cap:39K-sideband-cooling}Mass selective cooling
of $^{39}$K: \emph{a}) no excitation is applied \emph{b})
$\nu_{-}$ excitation; all peaks disappear \emph{c}) quadrupole
$\nu_{c}(^{39}\textnormal{K})$ excitation; $^{39}$K$^+$ ions come
back but the signal amplitude is, for some as yet unknown reason,
much lower.}
\end{figure}

A mass selective removal of unwanted species can be achieved via a
combination of dipole excitation of the magnetron motion
($\nu_{-}$) and quadrupole excitation at the true cyclotron
frequency ($\nu_{c}$)~\cite{savard91}. This technique was tried
qualitatively for $^{39}$K$^+$ ions. A dipole $\nu_{-}$ excitation
(at $\nu_{-}=130$~Hz, determined in a similar way as $\nu_{+}$,
i.e. checking at which frequency all ion peaks disappear from the
oscilloscope spectrum) with amplitude $A_{\nu_{-}}=150$~mV was
applied for 50\emph{~}ms, and followed by a quadrupole $\nu_{c}$
excitation ($\nu_{c}(^{39}\textnormal{K})=3553729$~Hz) with
amplitude $A_{\nu_{c}}(^{39}\textnormal{K})=1.6$~V for 3\emph{~}ms
(B-field is 9\emph{~}T). As in the previous tests no systematic
scan of the excitation frequencies was performed but only a visual
analysis of the TOF oscilloscope spectrum. The corresponding steps
of the process are shown in Fig.~\ref{cap:39K-sideband-cooling}.
Fig.~\ref{cap:39K-sideband-cooling}\emph{a} displays the situation
before any excitation. The second peak corresponds to
$^{39}$K$^+$. When a dipole $\nu_{-}$ excitation is applied, all
ions are driven out as can be seen in
Fig.~\ref{cap:39K-sideband-cooling}\emph{b} (both peaks
disappear). If now a quadrupole RF-field at frequency
$\nu_{c}(^{39}\textnormal{K})$ is used, one can expect the
$^{39}$K$^+$ ions to recenter while other impurities should
disappear. However, while it is clear from
Fig.~\ref{cap:39K-sideband-cooling}\emph{c} that there is indeed
no other species present than $^{39}$K$^+$, the signal
corresponding to the $^{39}$K$^+$ ions is significantly smaller
and broader than the one without any excitation. A possible reason
for this can be either wrongly chosen parameters (so that the ions
hit the electrode and are lost) or some electronics problem (e.g.
electronic noise in some channels, a different capacitance of the
track: connection wires + electrodes, or imperfection of power
supplies used). Also, the behaviour of the ions might have been
influenced by space charge effects.

The dipole $\nu_{-}$ excitation was also applied while working
with $^{20}$Ne$^+$ ions and the frequency used there was
$\nu_{-}\simeq140$~Hz (to successfully remove all species; B-field
was 9\emph{~}T). \label{dipole-minus}Combining this value with the
one in the $^{39}$K test leads to $\nu_{-}=135(5)$~Hz. Using now
this value of $\nu_{-}$ and an estimation of the magnetic field in
the cooler trap center (see next Sec.~\ref{sub:Magnetic-field}),
one can deduce the trap characteristic parameter
$U_{0}/d^{2}=1.53(6)\cdot10^{4}$~V/$\textnormal{m}^{2}$ for WITCH.
For the ISOLTRAP~\cite{bollen96,raimbault97,lunney00} cooler trap,
which is very similar to the WITCH cooler trap, one has
$U_{0}/d^{2}=1.8\cdot10^{4}$~
V/$\textnormal{m}^{2}$~\cite{beck-d-97}.

\subsection{Magnetic field\label{sub:Magnetic-field}}

The necessity to know precisely the magnetic field at the trap
center is based on the following two factors: 1) the cyclotron
frequency, and therefore the centering and cooling of the ion
cloud as well as the mass-selectivity, are directly determined by
the value of the field, while 2) the response function of the
WITCH spectrometer also depends on it. Originally the magnetic
field map was provided by Oxford Instruments for both magnets
separately. However, this field map resulted from a
calculation/approximation based on measurements that were made at
the factory, prior to the delivery and installation of the system
at CERN. Also, inserting the traps structure with all the wiring
may change the field strength in the trap centers due to the
magnetic susceptibility of the materials used. Direct measurement
of the field with an NMR probe is hardly possible because of the
very difficult access to the area.

An elegant way to estimate the magnetic field is to use a known
isotope (i.e. with known mass $m$) and experimentally find the
proper cyclotron frequency $\nu_{c}$. This yields enough
information to determine the magnetic field $B$~\cite{konig95}.
From the excitation tests performed with a $^{39}$K$^+$ beam,
$\nu_{c}(^{39}\textnormal{K})$ was found via quadrupole excitation
of $^{39}$K$^+$ (direct determination, $\nu_{c}=3553729(2000)$~Hz)
but also via dipole excitation of $^{39}$K$^+$ (which gives only
$\nu_{+}=3554019(2000)$~Hz; in another test $\nu_{-}=135(5)$~Hz
was found, see p.\pageref{dipole-minus}). The estimated
uncertainty for $\nu_{c}$ and $\nu_{+}$ is based on two
measurements of $\nu_{+}(^{39}\textnormal{K})$ at 6\emph{~}T
field: $\nu_{+}^{(1)}(^{39}\textnormal{K})=2363170$~Hz and
$\nu_{+}^{(2)}(^{39}\textnormal{K})=2367100$~Hz
(Fig.\ref{cap:peak-indent}), which leads to
$\Delta\nu_{+}(^{39}\textnormal{K})\backsimeq2000$~Hz. The final
result for $\nu_{c}(^{39}\textnormal{K})$ is
$\bar{\nu}_{c}=3553900(1400)$~Hz, which corresponds to a magnetic
field in the center of the cooler trap
\textbf{$B_{cooler}=9.018(4)$~}T (according to the field map from
Oxford Instruments the field in the center of the cooler trap is
$9.0011\:\textnormal{T}$). The set field was 9\emph{~}T for the
lower magnet and 0.1\emph{~}T for the top magnet of the system
(the WITCH magnet system allows to set any possible combination of
the magnetic fields in the range $0<B_{max}\le9$~T and
$0<B_{min}\le0.2$~T).

\subsection{MCP regime\label{sub:MCP-regime}}

\begin{figure}[H]
\centering\includegraphics[%
  width=0.99\textwidth,
  keepaspectratio]{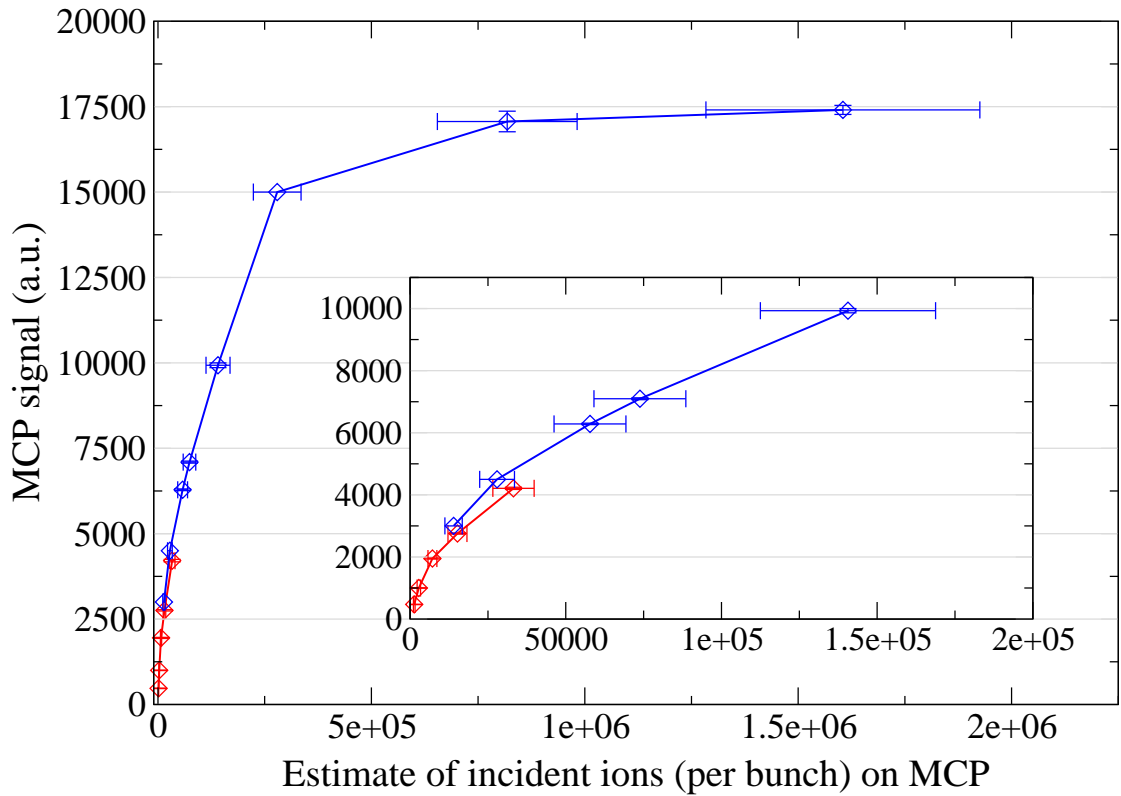}

\caption[Summed MCP signal as a function of the number of incident
particles (per bunch)]{\label{cap:mcp-nions}Summed MCP signal as a
function of the number of incident particles (per bunch) for MCP
HV = 1.4\emph{~}kV (third VBL diagnostics, the MCP sensitive area
is 18~mm diameter). The plot shows a combination of two
independent measurements. As can be seen, the MCP saturates if the
number of incident particles is $\gtrsim2\cdot10^{5}$ per bunch.}
\end{figure}

\begin{figure}
\centering\includegraphics[%
  width=0.9\textwidth,
  keepaspectratio]{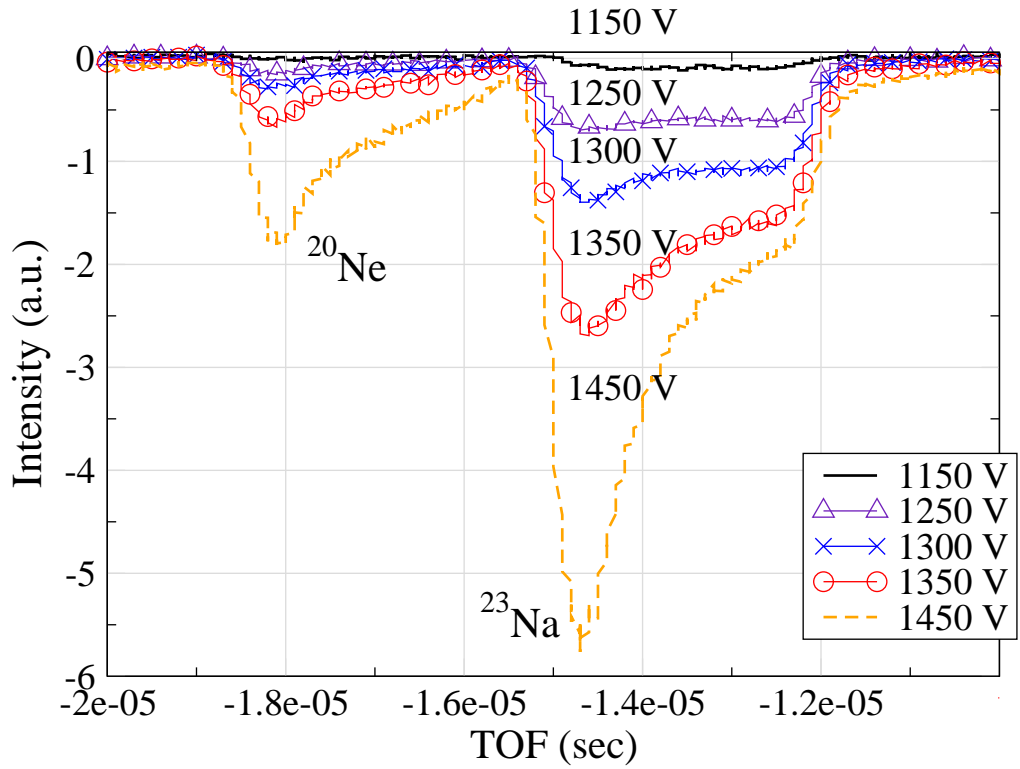}

\caption{\label{cap:mcp-hv}Change of MCP signal shape with applied
HV, and the saturation effect (measured with third VBL
diagnostics).}
\end{figure}

\begin{figure}[H]
\centering\includegraphics[%
  width=0.9\textwidth,
  keepaspectratio]{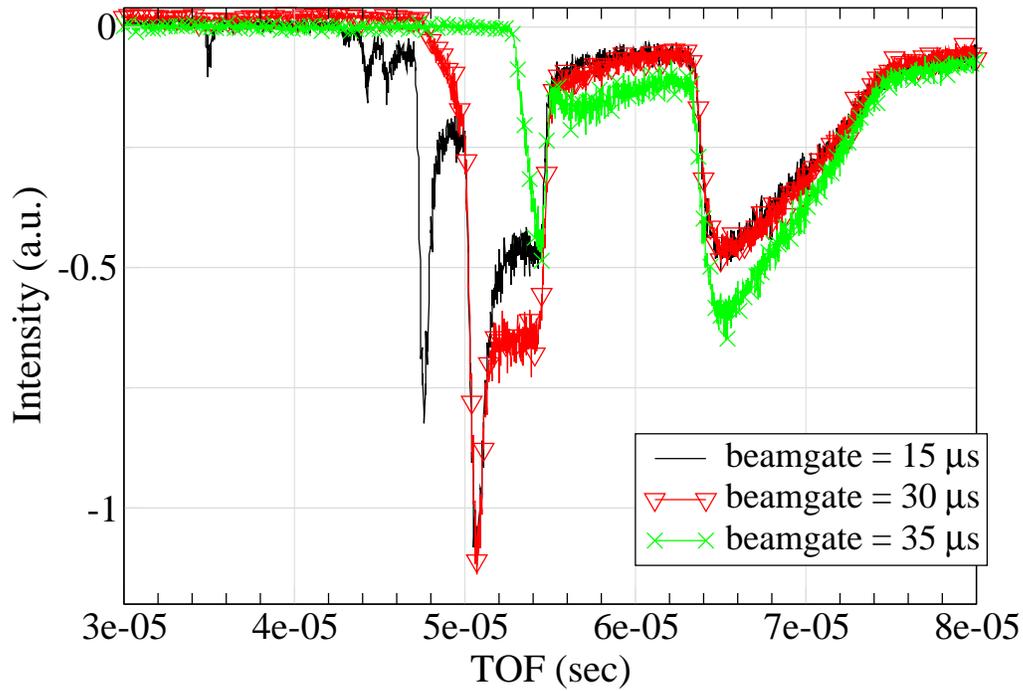}

\caption{\label{cap:mcp-beamgate}Effect of the WITCH beam gate on
the MCP signal (first VBL diagnostics, MCP HV=1.6\emph{~}kV):
removing early arriving ions with the beam gate increases the MCP
signal of later arriving ions.}
\end{figure}

The working principle of an MCP detector can be found in
e.g.~\cite{wiza79,gao84,brehm95}. Under certain conditions an MCP
is not sensitive anymore to the number of incident particles
(Fig.~\ref{cap:mcp-nions}), meaning that the detector misses some
events. This state is known as \emph{the saturation of the MCP}.
The typical \emph{dead-time} of one MCP channel is in the order of
several tens of milliseconds~\cite{wiza79}. For the WITCH
diagnostic MCPs it is $\sim$30\emph{~}ms. The saturation of an MCP
depends on both the ion current density (i.e. the number of ions
per MCP channel and per second) and the MCP acceleration voltage.
The effect of the ion current density is shown in
Fig.\ref{cap:mcp-nions}: when the number of ions in the bunch
exceeds a certain value the MCP signal remains constant. However,
already much earlier the dependence of the MCP signal on the
number of incident ions ceases to be linear. This behavior can be
explained by partial saturation: Fig.~\ref{cap:mcp-hv} shows that
at MCP HV=1.25\emph{~}kV \emph{}the $^{23}$Na$^{+}$ signal has a
block shape while at higher MCP voltages the total signal still
increases but shows a significant drop in intensity for later
arriving ions. This means that the early arriving ions saturate a
certain fraction of the MCP channels, leading to a decrease of the
MCP registration efficiency for later ions. This can also be seen
from Fig.~\ref{cap:mcp-beamgate}: when the early arriving ions are
removed with a time window before they reach the detector (the
WITCH beam gate is used for this), the signal corresponding to the
late ions increases.

The effect described above influences the measurements and has to
be taken into account for efficiency estimates and during the beam
tuning. Saturation of the diagnostic MCP's can be avoided by
adding several Ni meshes in front and using appropriate MCP
acceleration voltages. The dead-time and saturation of the MCP can
also influence the recoil spectrum measurement, since in real
measurement conditions the recoil ions are supposed to reach the
recoil MCP detector at a rate $>10^{5}$~Hz. Results of a careful
study of the MCP response to high intensity pulsed beams are
described in \cite{coeck05b}.

\subsection{First radioactive ions}

\begin{figure}[H]
\centering\includegraphics[%
  width=1.0\textwidth,
  keepaspectratio]{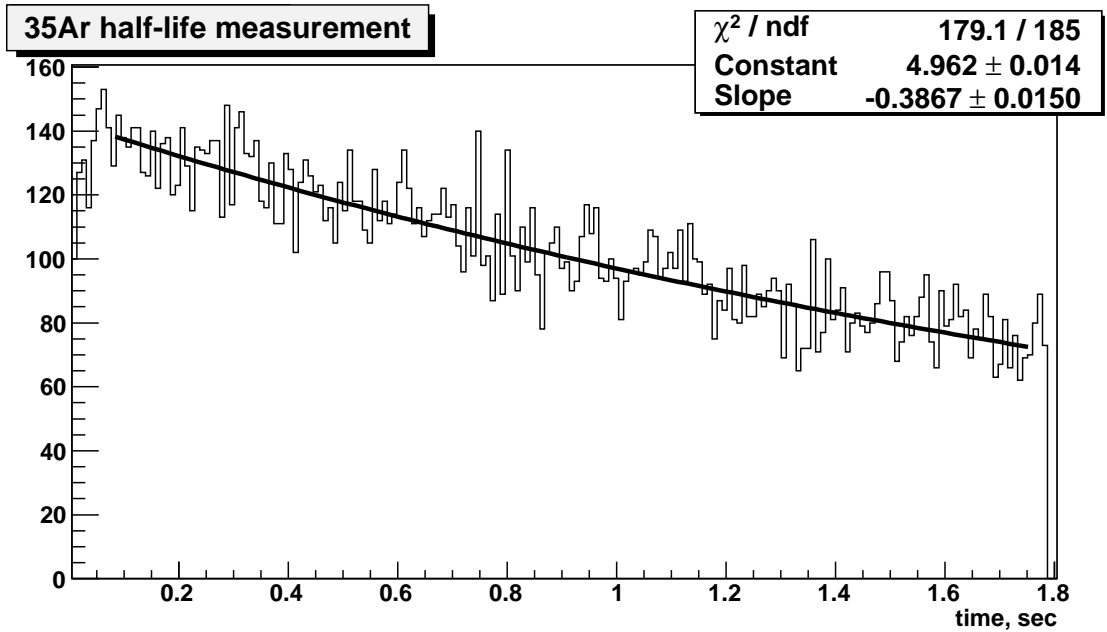}

\caption{\label{cap:35Ar-lifetime}$^{35}$Ar half-life measurement
on the first VBL diagnostic MCP detector.}
\end{figure}

In November 2004 WITCH got its first radioactive beam time (with
$^{35}$Ar$^+$). A \emph{}CaO ISOLDE target and a plasma ion source
with cold transfer line were used. The rate for this particular
run was about $5\cdot10^{5}\,\textnormal{atoms/s}$, being somewhat
lower then expected. Coupled with the fact that the efficiency of
WITCH is not yet high enough to deal with this
 intensity (see Table~\ref{cap:efficiency-witch}) a recoil
energy spectrum measurement was not yet possible. Increase of
ISOLDE rates of 40 times can be expected based on past
demonstrated yields \cite{isolde05a}. For WITCH ongoing
optimization should yield an increase in overall efficiency of a
factor of ten or more. Still, the decay of $^{35}$Ar was observed
on the first VBL MCP detector (Fig.~\ref{cap:35Ar-lifetime}). The
half life of $^{35}$Ar obtained as the weighted average of two
short measurements is: $T_{1/2}(^{35}\textnormal{Ar})=1.70(5)$~s
(the value in literature is 1.775(4)~s). This showed that ISOLDE
delivered a clean $^{35}$Ar$^+$ beam with no radioactive
contaminant, as is required for the planned recoil spectrum
measurements.

\section{Outlook \& improvements}

The WITCH set-up was completed and first commissioning tests performed
only recently. There is still room for many improvements and more
tests are necessary to better understand the behaviour of the different
parts of the set-up.

The non-pulsed high energy 60~keV (30\emph{~}keV) ions arrive
first on the diagnostic MCP and can cause saturation of the
detector, reducing its sensitivity, i.e. disturbing the tuning and
efficiency measurements. Part of these energetic ions also reaches
the detection MCP at the end of the spectrometer, in spite of the
magnetic field. This will influence the measurement of the recoil
ion spectrum. Another drawback is that during a radioactive run,
the decays of non-pulsed 60~keV (30\emph{~}keV) radioactive ions
implanted directly on the detection MCP will lead to additional
background. All these problems can be avoided if one uses a beam
gate installed in the HBL (or VBL) in order to select only the
part of the original beam corresponding to the correctly pulsed
ions. The required electronics to switch the voltages in the range
of 1000\emph{~}V within several 100\emph{~}ns is currently being
developed.

A new system of VBL diagnostics is currently being prepared. It is
based on an MCP detector with split anode and a Ni-mesh in front
of it. The latter reduces the intensity of the incoming beam in
order to avoid saturation of the detector. The transparency of
this mesh can be measured to good precision with laser light. The
split anode system provides the possibility to check the beam size
and its position {}``on-line''. The combination of the Ni-mesh and
the split anode will allow to avoid problems caused by the
saturation of the MCP during the beam tuning.

\begin{figure}[H]
\centering\includegraphics[%
  width=0.90\textwidth,
  keepaspectratio]{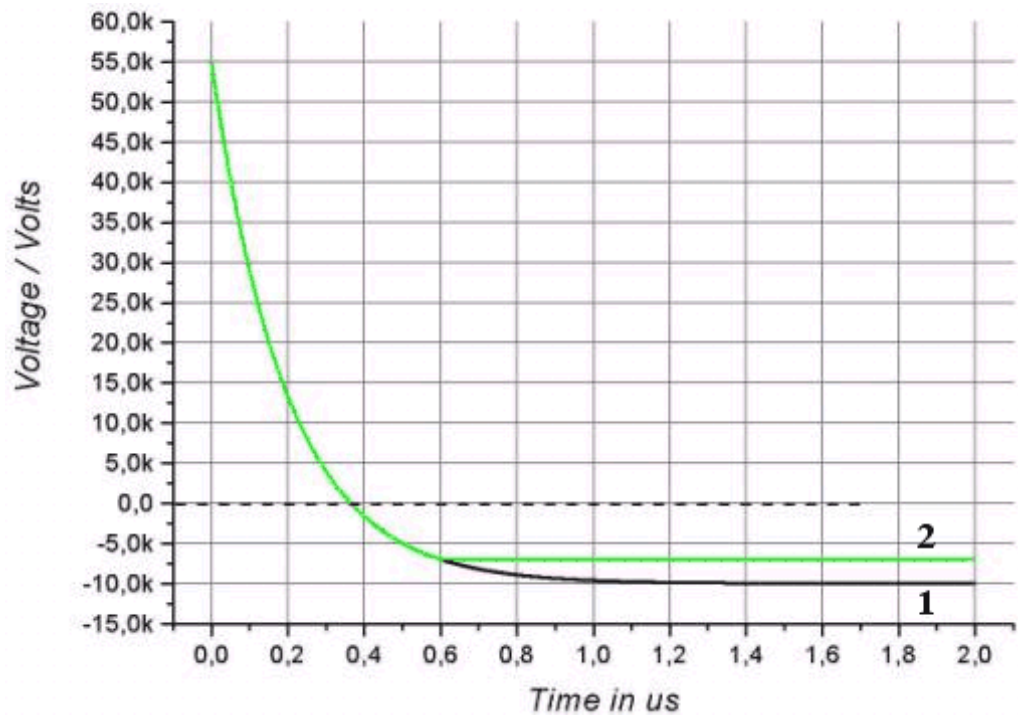}

\caption{\label{cap:hv-decay-new}HV of the PDT as a function of
time in case of a standard exponential decrease with
$\tau_{PDT}=0.2\,\mu s$ (1) and using the clamping diode (2) (from
\cite{stahl04}).}
\end{figure}

Based on the new HV switch system of ISOLTRAP a new HV scheme was
developed in close cooperation with a company%
\footnote{~Dr. Stefan Stahl - Elektronik-Beratung, Sonderanfertigungen · Kellerweg
23, D - 67582 Mettenheim · Germany.%
}. This new system is more reliable for 60~kV switching and has in
addition the advantage that the switching time is improved by a factor
of $2\div3$. This is done with an approach of a clamping diode, which
ties the decreasing voltage to a pre-defined value for a limited period
of time. The switching process in this case starts as usual (Fig.~\ref{cap:hv-decay-new}),
i.e. the voltage of the PDT goes down towards the negative biasing
voltage. After roughly 600\emph{~}ns, corresponding to $3\times\tau_{PDT}$,
the voltage has dropped below the value of an auxiliary voltage supply
to which the diode is connected. The diode therefore becomes conducting
and prevents the PDT-voltage from a further decrease.

The trap tests showed that the buffer gas of the cooler trap
contains impurities. The main effect of this is that the ions of
interest can be neutralized via charge exchange and can thus
escape from the trap. To avoid this, the external gas line has to
be as short as possible (the internal part cannot be changed),
while in addition it has to be cleaned, baked and pumped to remove
contaminations. To further clean the buffer gas one can in
addition install a cold trap or use a commercially available He
purifier. During the commissioning period it was realized that a
higher energy for the ions leaving the PDT improves the injection
in the cooler trap. However, this requires an upgrade of the end
cap power supplies. The corresponding electronics is currently
being developed. With respect to the detection part, an
8\emph{~}cm diameter MCP detector with position sensitive anode
will be used to study the size of the recoil ion beam, the
possible dependence of the beam size on the ion energy, as well as
the $\beta$-background. This MCP and the necessary electronics
will be provided by the LPC-Caen
group~\cite{roentdek05,lienard05}.

To carry out efficiency tests and improve the beam tuning in
WITCH, it is necessary to transport ion beams through the complete
WITCH beamline. The REXTRAP ion source is often needed by the
REXTRAP team for tests and, in addition, can not be used during
experiments involving the REXTRAP set-up (since this ion source
blocks the ISOLDE beam at the entrance of REXTRAP). A design study
was therefore started to develop an ion source for WITCH similar
to the one of REXTRAP and to implement this in the horizontal
beamline.

The vacuum of the WITCH system is at present reasonably good
($\sim10^{-8}\div10^{-7}$~mbar) for normal WITCH operation but can
still be improved to avoid pressure related systematic effects and
to reduce the charge exchange probability. The WITCH spectrometer
was designed with the possibility to use non-evaporable getters
(NEG) and it is planned to put this system in operation soon.

Finally, additional tests are being prepared in order to improve
the WITCH efficiency. For instance, one may try to optimize
REXTRAP operation so as to reduce the ion bunch length, try to
determine the rest gas pressure in the WITCH traps, study the size
of the ion cloud, investigate space charge effects, and try to
obtain an optimal cooling time. In addition, investigation of the
spectrometer and its response function, as well as measurements to
check the $\beta$-background on the main detector are planned as
well.

\section{Conclusion}

The installation period of the WITCH set-up, which was developed
over the last few years, was finished in autumn 2004 while
intensive commissioning of the set-up was performed during the
whole year 2004. The main aim of these tests was to check the
operation of the beam transport, the pulsing down of the ion beam
and the injection of ions into the high magnetic field, to test
the trap basics and check the spectrometer operation, and finally,
to optimize as many settings as possible. These tests showed that
the full set-up up to the spectrometer is now operational,
although several efficiencies still have to be improved. The
present overall efficiency of the experiment did not allow to
actively test the retardation spectrometer which can only be done
fully with radioactive ions. The results of the commissioning
stage were carefully analyzed and possible improvements were
suggested. This includes both technical modifications as well as
the necessary tests in order to optimize the set-up and achieve
the required efficiency.

\section*{Acknowledgement} This work is supported by the European
Union grants FMRX-CT97-0144 (the EUROTRAPS TMR network) and
HPRI-CT-2001-50034 (the NIPNET RTD network), by the Flemish Fund
for Scientific Research FWO and by the projects GOA 99-02 and GOA
2004/03 of the K.U.Leuven. This research was partly funded with a
specialization fellowship of the Flemish Institute for the
stimulation of Scientific-Technological Research in the Industry
(IWT). D.B. was supported by a Marie-Curie fellowship from the TMR
program of the European Union.

\newpage

\begin{table}
\centering\begin{tabular}{|p{7.5cm}|c|c|} \hline Description of
the parameter& \multicolumn{2}{c|}{Efficiency}\tabularnewline
\cline{2-3} & ideal set-up& achieved\tabularnewline \hline \hline
HBL efficiency, $\eta_{HBL}$& 100\%& $\sim$100\%\tabularnewline
\hline PDT efficiency, $\eta_{PDT}$& 43\%& 8\%\tabularnewline
\hline Injection into the magnetic field, $\epsilon_{injection}$ &
100\%& 1\%$\div$10\%\tabularnewline \hline Trapping in the cooler
trap& 100\%& $\sim$60\%\tabularnewline \hline Losses during
cooling& {\small \rule{0pt}{10pt}}100\% $^{a)}$&
$\sim75\%\;^{a),\, b),\, c)}$\tabularnewline \hline Efficiency of
transfer between traps& 100\%& $\sim80\%$\tabularnewline \hline
Losses in the decay trap& {\small \rule{0pt}{10pt}}100\% $^{a)}$&
{\small \rule{0pt}{10pt}}100\% $^{a),\, c)}$\tabularnewline \hline
Fraction of ions leaving the decay trap, taking into account the
cut-off angle ($\epsilon_{\theta}$) and solid angle
($\epsilon_{\Omega}$)& {\small \rule{0pt}{10pt}}40\% $^{d)}$& not
yet studied\tabularnewline \hline Shake-off for charge state
$n=1$, $p_{(n=1)}$& 10\%& not yet studied\tabularnewline \hline
Transmission through spectrometer& {\small \rule{0pt}{10pt}}100\%
$^{a)}$& not yet studied\tabularnewline \hline MCP efficiency,
$\epsilon_{MCP}$& 60\%& {\small \rule{0pt}{10pt}}52.3(3)\%
$^{e)}$\tabularnewline \hline \hline Total efficiency& $\sim1\%$&
{\small
\rule{0pt}{10pt}}$\sim0.6\cdot\left(10^{-3}\div10^{-2}\right)\%$$\;{}^{f)}$\tabularnewline
\hline
\multicolumn{3}{p{0.95\textwidth}}{{$^{a)}$~\rule{0pt}{11pt}\footnotesize
100\% means that there are no losses.}}\tabularnewline
\multicolumn{3}{p{0.95\textwidth}}{{$^{b)}$~\rule{0pt}{11pt}\footnotesize
after 300~ms of cooling, with 5~mbar He buffer gas pressure at the
measurement position.}}\tabularnewline
\multicolumn{3}{p{0.95\textwidth}}{{$^{c)}$~\rule{0pt}{11pt}\footnotesize
estimated from the tests performed, but one of the other effects
(e.g. MCP regime) was not taken into account.}}\tabularnewline
\multicolumn{3}{p{0.95\textwidth}}{{$^{d)}$~\rule{0pt}{11pt}\footnotesize
ion energy dependent value.}}\tabularnewline
\multicolumn{3}{p{0.95\textwidth}}{{$^{e)}$~\rule{0pt}{11pt}\footnotesize
from E. Li{\'e}nard et al., Nucl. Instr. and Meth. A, 551 (2005)
375.}}\tabularnewline
\multicolumn{3}{p{0.95\textwidth}}{{$^{f)}$~\rule{0pt}{11pt}\footnotesize
when an efficiency has not been determined yet the values for an
ideal set-up are taken.}}\tabularnewline
\end{tabular}

\caption{\label{cap:efficiency-witch}Efficiencies for a fully
optimized WITCH set-up and the currently achieved values.}
\end{table}


\begin{thebibliography}{10}
\bibitem{lee56}T.D. Lee and C.N. Yang, Phys. Rev., 104 (1956) 254.
\bibitem{jackson57a}J.D. Jackson and S.B. Treiman and H.W. Wyld, Phys. Rev., 106 (1957)
517.
\bibitem{severijns05a}N. Severijns and M. Beck and O. Naviliat-Cuncic, Rev. Mod. Phys.,
(2006) to be published.
\bibitem{jackson57b}J.D. Jackson and S.B. Treiman and H.W. Wyld, Nucl. Phys., 4 (1957)
206.
\bibitem{hardy05}J.C. Hardy and I.S. Towner, Phys. Rev. Lett., 94 (2005) 092502.
\bibitem{wu66}C.S. Wu and S.A. Moszkowski, Beta Decay, John Wiley \& Sons Inc., New
York, 1966.
\bibitem{brown86}L.S. Brown and G. Gabrielse, Rev. Mod. Phys., 58 (1986) 233.
\bibitem{beck03a}M. Beck et al., Nucl.Instr. and Meth. A, 503 (2003) 567.
\bibitem{picard92}A. Picard et al., Nucl. Instr. and Meth. B, 63 (1992) 345.
\bibitem{lobashev85}V.M. Lobashev and P.E. Spivak, Nucl. Instr. and Meth. A, 240 (1985)
305.
\bibitem{jackson75}J.D. Jackson, Classical Electrodynamics, John Wiley \& Sons Inc.,
New York, 1975
\bibitem{kozlov03}V.Yu. Kozlov, M. Beck et al., Phys. of Atom. Nucl., 67 (2004) 1112.
\bibitem{carlson63}T.A. Carlson, F. Pleasonton, and C.H. Johnson,
Phys. Rev., 129 (1963) 2220.
\bibitem{kugler92}E. Kugler et al., Nucl. Instr. and Meth. B, 70 (1992) 41.
\bibitem{kugler00}E. Kugler, Hyp. Int., 129 (2000) 23.
\bibitem{isolde05a}ISOLDE Collaboration, http://isolde.web.cern.ch/ISOLDE/
\bibitem{ames05}F. Ames et al., Nucl. Instr. and Meth. A, 538 (2005) 17.
\bibitem{herfurth01b}F. Herfurth et al., Nucl. Instr. and Meth. A, 469 (2001) 254.
\bibitem{beck-d-02}D. Beck et al., Nucl. Phys. A, 701 (2002) 369.
\bibitem{phd-delaure}B. Delaur{\'e}, Development of a Penning trap based Set-up for
Precision Tests of the Standard Model, Ph.D. thesis, Katholieke
Universiteit Leuven, 2004
\bibitem{gorelov00}A. Gorelov et al., Hyp. Int., 127 (2000) 373.
\bibitem{snell68}A.H. Snell, The Atomic and Molecular Consequences of Radioactive Decay,
in Alpha-, Beta- and Gamma-Ray Spectroscopy, ed. Siegbahn, North
Holland, 1968, 1545.
\bibitem{snell55}A.H. Snell and F. Pleasonton, Phys. Rev., 100 (1955) 1396.
\bibitem{scielzo03}N. Scielzo et al., Phys. Rev. A, 68 (2003) 022716.
\bibitem{geant4-03}S. Agostinelli et al., Nucl. Instr. and Meth. A, 506 (2003) 250.
\bibitem{geant4-web}GEANT4 Home page, http://geant4.web.cern.ch/geant4
\bibitem{phd-kozlov}V.Yu. Kozlov, WITCH, a Penning trap for weak interaction
studies, Ph.D. thesis, Katholieke Universiteit Leuven, 2005.
CERN-THESIS-2006-009
\bibitem{delaure05}B. Delaur{\'e} et al., Nucl. Instr. and Meth. A, to be published.
\bibitem{dahl95}D.A. Dahl, SIMION 3D 6.0: users manual, Princeton Electronic systems,
Inc., 1995.
\bibitem{savard91}G. Savard et al., Phys. Lett. A, 158 (1991) 247.
\bibitem{bollen96}G. Bollen et al., Nucl. Instr. and Meth. A, 368 (1996) 675.
\bibitem{raimbault97}H. Raimbault-Hartman et al., Nucl. Instr. and Meth. B, 126 (1997)
378.
\bibitem{lunney00}D. Lunney and G.Bollen, Hyp. Int., 129 (2000) 249.
\bibitem{beck-d-97}D. Beck, Ph.D. thesis, Johannes Gutenberg Universit{\"a}t Mainz,
1997.
\bibitem{konig95}M. K{\"o}nig et al., Int. J. Mass Spec., 142 (1995) 95.
\bibitem{wiza79}J.L. Wiza, Nucl. Instr. and Meth., 162 (1979) 587.
\bibitem{gao84}R.S. Gao et al., Rev. Sci. Instr., 55 (1984) 1756.
\bibitem{brehm95}B. Brehm et al., Meas. Sci. Tech., 6 (1995) 953.
\bibitem{coeck05b}S. Coeck et al., Nucl. Instr. Meth. A, 557 (2006) 516.
\bibitem{stahl04}S. Stahl, Technical report, Elektronik-Beratung, Sonderanfertigungen,
Kellerweg 23, D - 67582 Mettenheim · Germany (2004).
\bibitem{roentdek05}Roentdek Handels GmbH, Kelkheim-Ruppertshain, http://www.roentdek.de
\bibitem{lienard05}E. Li{\'e}nard et al., Nucl. Instr. and Meth. A, 551 (2005) 375.
\bibitem{behr05}J. Behr, private communication.
\bibitem{scielzo05}N. Scielzo, private communication.
\end{thebibliography}
\end{document}